\documentclass[a4paper,11pt]{article}
\pdfoutput=1
\usepackage{jcappub}%[=2023-09-11]
\keywords{inflation, primordial black holes, cosmological perturbation theory}
\arxivnumber{2303.12025}
\newcommand{\rdddmathspace}{\thinmuskip=0mu%
\medmuskip=0mu%
\thickmuskip=1mu plus 3mu}

\newcommand{\calR}{{\cal{R}}}
\newcommand{\calP}{{\cal{P}}}

\newcommand{\bfq}{{\bf{q}}}
\newcommand{\bfp}{{\bf{p}}}
\newcommand{\Ha}{{\bf H}_3  }
\newcommand{\Hb}{{\bf H}_4  }

\title{One-loop corrections in  power spectrum in single field inflation}

\author{Hassan Firouzjahi}

\affiliation{School of Astronomy, Institute for Research in Fundamental Sciences (IPM),\\
P.O.~Box 19395-5531, Tehran, Iran}

\emailAdd{firouz@ipm.ir}

\abstract{\looseness=-1 We revisit the one-loop correction in curvature perturbation power spectrum in models of  single field  inflation which undergo a phase of ultra slow-roll (USR) inflation. We include the contributions from both the cubic and quartic interaction Hamiltonians and calculate the one-loop corrections on the spectrum of the CMB scale modes from the small scale modes which leave the horizon during the USR phase. It is shown that the amplitude of one-loop corrections depends on the sharpness of the transition from the USR phase  to the final slow-roll phase. For an arbitrarily sharp transition, the one-loop correction becomes arbitrarily large, invalidating the perturbative treatment of the analysis. We speculate that for a mild transition, the large one-loop corrections are washed out  during the subsequent evolution after the USR phase. The implications for primordial black holes formation are briefly reviewed.}

\RequirePackage{expl3}
\RequirePackage{xparse}
\ExplSyntaxOn
\cs_new:Nn \l_https_cs:nn {
    \href{https\char_generate:nn{58}{12}//#1}{#2}
}
\cs_new:Nn \l_https_arx_cs:nn {
    \l_https_cs:nn {arxiv.org/abs/#1} {#2}
}
\cs_new:Nn \l_https_doi_cs:nn {
    \l_https_cs:nn {doi.org/#1} {#2}
}
\cs_new:Nn \l_https_ins_cs:nn {
    \l_https_cs:nn {inspirehep.net/#1} {#2}
}
\NewDocumentCommand{\uniarXiv}{m}{
    \regex_match:nnTF { [a-z]{2,} } {#1}
    {
        \l_https_arx_cs:nn {#1} {\texttt{#1}}
    }
    {
        \l_https_arx_cs:nn {#1} {\texttt{arXiv:#1}}
    }
}
\NewDocumentCommand{\unidoi}{m}{
    \tl_set:Nn \l_link_tl {#1}
    \tl_set:Nn \l_text_tl {#1}
    \regex_replace_all:nnN { _ } { \cL_ } \l_text_tl
    \l_https_doi_cs:nn {\l_link_tl} {\texttt{\footnotesize DOI:\l_text_tl}}
}
\NewDocumentCommand{\uninspire}{m}{
    \l_https_ins_cs:nn {#1} {{\tiny IN}{\footnotesize SPIRE}}
}
\ExplSyntaxOff

\newcommand\erratum[4][ibid.\ ]{\emph{Erratum #1}{\bf #2} (#3) #4}

\begin{document}
\maketitle\flushbottom

\section{Introduction}
\label{intro}

The inflationary Universe is a successful paradigm for early universe cosmology
which is well supported by cosmological observations~\cite{Planck-ml-2018vyg, Planck-ml-2018jri}.  In the simplest scenario, inflation is driven by a single scalar field which slowly rolls on top of a nearly flat potential. Among the basic predictions of models of inflation are that the primordial perturbations are nearly scale invariant, Gaussian and adiabatic.

\looseness=-1
A key prediction of most single field models of inflation is that the curvature perturbations on superhorizon scales are constant~\cite{Kodama-ml-1984ziu, Mukhanov-ml-1990me, Weinberg-ml-2008zzc}. The conservation of curvature perturbation on superhorizon scales has the great advantage that one can read off the fundamental properties of the primordial  inflationary universe at the late  time when the corresponding modes  reenters the horizon. Because of this fundamental property, single field models of inflation are particularly interesting. For example, it was shown by Maldacena~\cite{Maldacena-ml-2002vr} that there is a consistency relation~\cite{Creminelli-ml-2004yq} between the amplitude of non-Gaussianity parameter $f_{NL}$ in the squeezed limit  and the spectral index of curvature perturbation $n_s-1$, for a review see~\cite{Chen-ml-2010xka, Komatsu-ml-2010hc}.  While this consistency condition applies to a vast majority of single field models, however it has a well-known
counter example, the ultra slow-roll (USR) setup, in which the consistency condition is violated~\cite{Namjoo-ml-2012aa}. USR setup is a model in which the potential is very flat so the inflaton speed falls off exponentially during inflation~\cite{Kinney-ml-2005vj, Martin-ml-2012pe, Morse-ml-2018kda, Lin-ml-2019fcz}. Consequently the curvature perturbations grow on superhorizon scale which is the main reason behind the violation of the consistency condition~\cite{Chen-ml-2013aj, Akhshik-ml-2015rwa, Mooij-ml-2015yka, Bravo-ml-2017wyw, Finelli-ml-2017fml}.
In particular, it is shown in~\cite{Namjoo-ml-2012aa}
that local-type non-Gaussianity with the amplitude $f_{NL}= \frac{5}{2}$ is generated in USR setup with a canonical kinetic energy.   This value can be enhanced if one considers models of non-attractor inflation with non-trivial sound speed~\cite{Chen-ml-2013aj, Chen-ml-2013eea}.

\looseness=-1
There have been significant interests in primordial black holes (PBHs) formation in the setup of USR inflation~\cite{Ivanov-ml-1994pa, Garcia-Bellido-ml-2017mdw, Biagetti-ml-2018pjj}, for a recent review see~\cite{Ozsoy-ml-2023ryl} and references therein.  In order for the PBHs to comprise a sizeable fraction of dark matter energy density, one requires that the amplitude of curvature perturbations to grow by about $10^7$ orders of magnitude
compared to its amplitude in CMB scale. This can be achieved easily if one considers a short period of USR inflation between two extended periods of slow-roll inflation. However, the viability of this setup for PBHs formation
was questioned recently in~\cite{Kristiano-ml-2022maq}  and~\cite{Kristiano-ml-2023scm}.
More specifically,  it was argued in~\cite{Kristiano-ml-2022maq, Kristiano-ml-2023scm}
that the one-loop corrections induced from small scale perturbations which leave the horizon during the USR phase will induce large contributions in curvature perturbation power spectrum, invalidating the perturbation theory. This is somewhat counterintuitive because one expects the superhorizon modes to be largely unaffected by small scale modes which leave the horizon only much later during the USR phase. However, following the example of the violation of consistency condition, things are a bit non-trivial in USR setup compared to usual slow-roll model. The would-be decaying mode in USR phase  is actually the growing mode  and a large growth of curvature perturbation can induce a source term for the large scale curvature perturbation which may not support the usual intuition. In addition, a key role is played by the mechanism in which the USR phase is terminated to be followed by a final attractor slow-roll phase. The danger of large one-loop effects may seem to be more sever for sharp transitions while for mild transitions the large one-loop corrections may be washed out during the transition from the USR phase to the final attractor slow-roll phase. In this work we revisit the question of one-loop correction in power spectrum induced from small scale modes which leave the horizon during the USR phase.  To study the system analytically  we consider a general sharp transition and compare our results with those of~\cite{Kristiano-ml-2022maq, Kristiano-ml-2023scm}.

The rest of the paper is organized as follows. In section~\ref{setup} we present our setup with emphasis on the mechanism of sharp transition from a USR phase to a slow-roll phase. In section~\ref{eft-action} we construct the cubic and quartic Hamiltonians using the formalism of effective field theory (EFT) of inflation. As a non-trivial application,  in section~\ref{bispectrum} we employ the cubic action obtained from our EFT approach and calculate the bispectrum, confirming the previous results in literature.  The one-loop corrections in power spectrum are presented in details in section~\ref{one-loop}
followed by summary and discussions in section~\ref{Summary}. Some technicalities concerning the in-in integrals of cubic Hamiltonian are relegated to appendix~\ref{appendix}.

\section{USR inflation setup}
\label{setup}

In this section we review our setup and gather the equations necessary for our follow up analysis.

As reviewed in previous section, the USR setup is among the very few models of single field inflation which violates the Maldacena consistency condition~\cite{Maldacena-ml-2002vr} relating the amplitude of non-Gaussianity parameter $f_{NL}$ in the squeezed limit to the spectral index $n_s-1$~\cite{Namjoo-ml-2012aa, Chen-ml-2013aj, Chen-ml-2013eea}.
The origin of the violation of the consistency condition is that the curvature perturbation is not frozen on superhorizon scale. In the simple picture where a USR setup is abruptly terminated by a slow-roll phase it was shown in~\cite{Namjoo-ml-2012aa} that $f_{NL}=\frac{5}{2}$. However, as demonstrated in~\cite{Cai-ml-2018dkf},
the final value of $f_{NL}$ at the end of inflation depends on the details of transition from the USR phase to the slow-roll phase. If the transition happens mildly then much of the  value of $f_{NL}$  is washed out during the subsequent evolution of the curvature perturbation in the final slow-roll phase. Consequently,  one obtains $f_{NL}$ at the order of the slow-roll parameter at the end of inflation, though the consistency condition is still violated.

Recently the question of one-loop correction in USR setup has
attracted interests~\cite{Kristiano-ml-2022maq, Kristiano-ml-2023scm, Riotto-ml-2023gpm, Riotto-ml-2023hoz} (for an earlier work see~\cite{Cheng-ml-2021lif}).
More specifically, it was argued in~\cite{Kristiano-ml-2022maq}
and~\cite{Kristiano-ml-2023scm} that the one-loop corrections from the modes which
leave the Hubble radius during the USR phase can affect the large CMB scale modes such that the induced power spectrum become comparable to the tree-level power spectrum. This would inevitably violate the perturbative nature of the system
and causes concern. In addition, it was shown that this large one-loop correction prohibits the formation of PBHs in USR setup. These conclusions were criticized
in~\cite{Riotto-ml-2023gpm, Riotto-ml-2023hoz}. In particular, it was argued on physical grounds  in~\cite{Riotto-ml-2023hoz} that if one considers a physical situation where the transition from the USR phase to the final slow-roll phase takes place mildly, then the dangerous one-loop corrections are washed out and the mechanism of PBHs formation in USR setup is still viable. We  comment that the one-loop correction in power spectrum was also studied in~\cite{Inomata-ml-2022yte} in a model of inflation where perturbations are resonantly amplified because of the oscillatory features  in potential.  It was shown that the one-loop correction can become large to invalidate the perturbative assumption.

In this work we revisit these issues in more details, building upon the analysis of~\cite{Kristiano-ml-2022maq, Kristiano-ml-2023scm}.  As we see the mechanism of transition from the USR phase to the final slow-roll phase plays important roles so here we briefly review some analysis concerning the transition as studied in more details in~\cite{Cai-ml-2018dkf}. To study the evolution analytically, below we consider a sharp transition from a USR phase to the slow-roll phase  while a mild transition requires further numerical investigation which is beyond the scope of the current work. Also we consider a scalar field with a canonical kinetic energy with the sound speed $c_s=1$.

 The picture we have in  mind is a single field inflation driven by the scalar field $\phi$ with the potential $V(\phi)$. The potential is constructed such that we have three distinct phases of inflation. The first phase is the usual slow-roll inflation in which the inflaton slowly rolls on top of the potential. We assume that this period is long enough, say about 20-30 e-folds such that the observed CMB and large scale structure perturbations are generated in this phase. The curvature perturbation is nearly scale invariant with an amplitude fixed by the COBE normalization. The second stage is the USR phase in which the potential becomes exactly flat such that the inflaton velocity falls off exponentially and the curvature perturbation grows exponentially. This is the phase which was employed extensively in the literature
 to generate PBHs formation from the enhanced curvature perturbations. Of course, one has to terminate the USR phase so the curvature perturbation does not grow very large to invalidate the perturbation theory. Therefore, the final stage of inflation is again a slow-roll phase following the intermediate USR phase. Typically, one requires the duration of the USR phase to be about a few e-folds to obtain a sizeable fraction of the dark matter.

With the FLRW metric
\begin{equation}
ds^2 = -dt^2 + a(t)^2 d{\bf x}^2 \, ,
\end{equation}
the evolution of the system in the USR phase is given by
\begin{equation}
\ddot \phi(t) + 3 H \dot \phi(t)=0\, , \quad \quad 3 M_P^2 H^2 \simeq V_0,
\end{equation}
in which $M_P$ is the reduced Planck mass, $H$ is the Hubble expansion rate during inflation and $V_0$ is the value of the potential during the USR phase. As $V_0$ is constant, $H$ is very nearly constant while
$\dot \phi \propto \frac{1}{a^3}$. To parameterize the slow-roll dynamics of the inflaton field, let us define the two slow-roll  parameters related to  $H$  as follows,
\begin{equation}
\label{ep-eta}
\epsilon \equiv -\frac{\dot H}{H^2} =\frac{\dot \phi^2}{2 M_P^2 H^2}\, , \quad \quad
\eta \equiv \frac{\dot \epsilon}{H \epsilon} \, . \end{equation}
During the usual slow-roll phase both $\epsilon$ and $\eta$ are nearly constant and are small, say at the order $10^{-2}$.  However, during the USR phase, $\epsilon$ falls off like $a^{-6}$
while $\eta\simeq -6$. This is the hallmark of the USR inflation~\cite{Kinney-ml-2005vj}.
 Going to conformal time $d \tau= dt/a(t)$ in which $a H \tau \simeq -1$, we can parameterize $\epsilon(\tau)$ as
 \begin{equation}
 \epsilon(\tau) = \epsilon_i \bigg( \frac{\tau}{\tau_i} \bigg)^6 \, ,
 \end{equation}
in which $\epsilon_i$ is the value of $\epsilon$ at the start of inflation and prior to USR phase. The USR phase is during the period $\tau_i < \tau <\tau_e$ so the value of $\epsilon$ at the end of USR phase is $\epsilon_e = \epsilon_i \big( \frac{\tau_e}{\tau_i} \big)^6 $. Defining the number of e-fold as $d N= H dt$, the duration of the USR phase is denoted by $\Delta N \equiv N(\tau_e) - N(\tau_i)$ and
$\epsilon_e = e^{-6 \Delta N} \epsilon_i $.

Following the assumption of~\cite{Cai-ml-2018dkf}, suppose the potential after the USR phase are in the form to support a period of slow-roll inflation such that
\begin{equation}
V(\phi) = V(\phi_e) + \sqrt{2 \epsilon_V} V(\phi_e)  (\phi -\phi_e) + \frac{\eta_V}{2} V(\phi_e) (\phi -\phi_e)^2 + \ldots  \, .
\end{equation}
Here $2\epsilon_V \equiv \big(V'(\phi_e)/V(\phi_e) \big)^2$ and $\eta_V\equiv V''(\phi_e)/V(\phi_e)$ are the  slow-roll parameters associated to the
potential in the third slow-roll phase. From the above expansion we see that the potential is continuous at $\phi=\phi_e$. If one further requires that the derivative of the potential to be continuous as well then $\epsilon_V=0$ and the transition becomes smooth. On the other hand, if $\epsilon_V \neq 0$, then the derivative of the potential is not continuous and there is a kink in the potential. Depending on the ratio $\frac{\epsilon_V}{\eta_V}$ the transition can be either mild or sharp. As we are interested in a sharp transition  to handle  the analysis analytically, below we consider $\eta_V=0$. However, this is not a restrictive assumption and most of our analysis will be carried out to the case where $\eta_V=0$ as long as $\epsilon_V > | \eta_V|$ to guarantee a sharp transition.

 The background field equation is given by~\cite{Cai-ml-2018dkf}
 \begin{equation}
 \frac{d^2 \phi}{ d N^2} + 3 \frac{d \phi}{d N} + 3 M_P \sqrt{2 \epsilon_V} \simeq 0 \, ,
 \quad \quad 3 M_P^2 H^2 \simeq V(\phi_e) \, .
 \end{equation}
The above equation can be solved easily. Without loss of generality let us assume
the time of transition corresponds to $N=0$.  Imposing the continuity of $\phi$ and $\frac{d \phi}{d N}$  at $N=0$, we obtain
\begin{equation}
M_P^{-1}\phi(N)= \frac{C_1}{3} e^{-3 N} + \frac{h}{6}  \sqrt{2 \epsilon_V} N + C_2 \, ,
\end{equation}
 in which the constants of integration $C_1$ and $C_2$ are obtained to be
 \begin{equation}
 C_1=  \sqrt{2 \epsilon_e} \left(1 + \frac{h}{6} \right) \, , \quad \quad
 C_2 = M_P^{-1} \phi_e - \frac{ \sqrt{2 \epsilon_e}}{3} \left(1 + \frac{h}{6} \right) \, .
 \end{equation}
Following~\cite{Cai-ml-2018dkf}  we have defined the parameter $h$ as
\begin{equation}
\label{h-def}
h\equiv \frac{6 \sqrt{2 \epsilon_V} }{\dot \phi(t_e)} M_P = -6 \sqrt{\frac{\epsilon_V}{\epsilon_e}} \, .
\end{equation}
Note that since we assume that $\phi$ is decreasing during its evolution, then $\dot \phi<0$ so $h<0$.  As we shall see, $h$ is the key parameter of our model controlling the sharpness of the transition from the USR phase to the third slow-roll phase.

Finally, the slow-roll parameter, as defined via eq.~(\ref{ep-eta}) in the final slow-roll phase $(N>0)$ are given by
\begin{equation}
\label{ep-N}
\epsilon(\tau)= \epsilon_e  \bigg(\frac{h}{6} - \left(1+ \frac{h}{6} \right) \bigg(\frac{\tau}{\tau_e} \bigg)^3 \bigg)^{2}
\end{equation}
and
\begin{equation}
\label{eta-N}
\eta(\tau) = -\frac{6 (6+h)}{(6+h) - h   \big(\frac{\tau_e}{\tau} \big)^3} \, .
\end{equation}
From the above formulas we see that towards the final stage of inflation, $\tau \rightarrow \tau_0 \rightarrow 0$, $\epsilon \rightarrow \epsilon_e (\frac{h}{6})^2$ while $\eta $ vanishes like $\tau^3$. While $\epsilon$ is by construction  smooth  at the transition point but $\eta$ has a radical change at $\tau=\tau_e$. Right before the transition (i.e.\ during the USR phase) $\eta=-6$
while right after transition  $\eta= -6-h$. Therefore, very near the transition we can approximate $\eta$ as follows~\cite{Cai-ml-2018dkf}
\begin{equation}
\eta = -6 - h \theta(\tau -\tau_e) \quad \quad  \tau_e^- < \tau < \tau_e^+ \, .
\end{equation}
In particular, from the above approximation, we have
\begin{equation}
\label{eta-jump}
\frac{d \eta}{d \tau} = - h \delta (\tau -\tau_e)  \, ,  \quad \quad  \tau_e^- < \tau < \tau_e^+ \, .
\end{equation}
We consider two cases of sharp transition: ``natural'' sharp transition in which
$\eta$ drops to zero right after transition corresponding to $h=-6$. In this case
$\epsilon$ after the transition is frozen to its value at the end of USR, $\epsilon_e$.
This is the limit which was studied in~\cite{Kristiano-ml-2022maq, Kristiano-ml-2023scm}.
The other case is ``extreme'' sharp transition in which $|h| \gg 1$. In this case, $\epsilon$ after the transition evolves to a larger value and at the end of inflation (or when the evolution in final stage has reached its attractor phase) we have the final value  $\epsilon(\tau_0) \simeq \epsilon_V = \epsilon_e (\frac{h}{6})^2$.

\begin{figure}
	\centering
	\includegraphics[ width=0.68\linewidth]{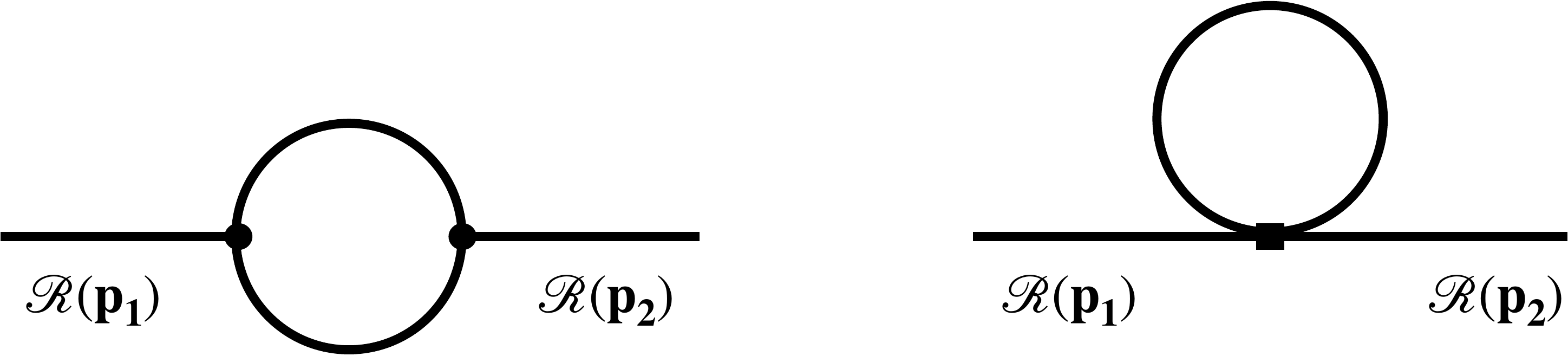}
	\caption{The Feynman diagrams for the one-loop correction in curvature perturbation power spectrum $\langle \calR(\bfp_1) \calR(\bfp_2) \rangle$.
The left and right panel represent the contribution  of the cubic Hamiltonian and
quartic Hamiltonian respectively.  }
\label{Feynman-fig}
\end{figure}

\section{Cubic and quartic action in EFT approach}
\label{eft-action}

\looseness=-1
 We are interested in one-loop corrections in power spectrum $\langle \calR(\bfp_1) \calR(\bfp_2) \rangle $ in which $\calR$ is the  comoving curvature perturbation. To calculate the one-loop corrections
one has to consider two different types of Feynman diagram as illustrated in figure~\ref{Feynman-fig}. The left panel corresponds to corrections induced from a cubic self interaction Hamiltonian. To obtain the cubic Hamiltonian, one has to calculate the action of curvature perturbation  to third order.
This was performed in details in~\cite{Maldacena-ml-2002vr}. However, to calculate the one-loop correction from the right panel of figure~\ref{Feynman-fig} one has to consider the quartic self interaction Hamiltonian. To calculate the quartic Hamiltonian one has to calculate the  action to fourth order for the perturbations. This is a difficult task and requires extra cares as there are contributions from the first and higher derivatives of $\eta$ which undergo jumps at $\tau=\tau_e$. It was argued briefly  in~\cite{Kristiano-ml-2022maq} that the quartic self interaction does
not contribute as they are suppressed by higher powers of $\epsilon$ so the authors of~\cite{Kristiano-ml-2022maq}  concentrated  only on the one-loop correction from the diagram in the left panel of figure~\ref{Feynman-fig}. Below, we shall show that this assumption is not correct and indeed there are significant contributions from the quartic self-interaction associated to  the right panel of figure~\ref{Feynman-fig}.

To calculate the cubic and quartic actions of curvature perturbations,  we employ the effective field theory (EFT) of inflation~\cite{Cheung-ml-2007st, Cheung-ml-2007sv}. This is a particularly useful approach in our setup where we are only interested in leading corrections and are not concerned with the sub-leading slow-roll corrections in interaction Hamiltonians. It is well-known that EFT of inflation is powerful in the decoupling limit where one neglects the gravitational back-reactions when only the matter perturbations are the dominant contributions.   Here we follow the method employed in~\cite{Akhshik-ml-2015nfa} which has used the method of EFT of inflation   to calculate the bispectrum for a general $P(X, \phi)$-type of non-attractor setup. We comment that  the EFT approach was employed recently~\cite{Choudhury-ml-2023jlt, Choudhury-ml-2023rks} as well to calculate the loop corrections in curvature perturbation power spectrum. However, similar to~\cite{Kristiano-ml-2022maq} and~\cite{Choudhury-ml-2023vuj}, these works also neglected the contributions associated to the Feynman diagram in the right panel of figure~\ref{Feynman-fig}.

In a near dS background with a time-dependent inflaton field $\phi(t)$, the four-dimensional diffeomorphism invariance  is spontaneously broken to a three-dimensional spatial diffeomorphism invariance. In  the unitary (or comoving) gauge
where the perturbations of inflaton are turned off
one  is allowed to write down all terms in the action which are consistent with the remaining three-dimensional diffeomorphsim invariance. In doing so,  the background inflation dynamics is controlled via the  known  Hubble expansion rate $H(t)$ and its derivative $\dot H(t)$. After writing the full action consistent with the three dimensional diffeomorphsim invariance,  one can restore the full four-dimensional diffeomorphsim invariance  by introducing a scalar field fluctuations, $\pi(x^{\mu})$, the Goldstone boson associated with the breaking of  the time diffeomorphsim invariance. One big advantage of the EFT approach is when one goes to the decoupling limit where the gravitational back-reactions are neglected. This corresponds to neglecting the slow-roll suppressed interactions in cubic and quartic actions while keeping only the leading terms which are relevant for large non-Gaussianities. Indeed, these interactions are those which induce large corrections in one loop integrals as well.

The full action is $S_{\rm total }= S_{\rm matter}+ S_{\rm EH}$ in which $S_{\rm matter}$ is the matter part of the action while $ S_{\rm EH}$ represents the usual Einstein-Hilbert action. In the decoupling limit, we do not perturb gravitational action and all leading perturbations come from the matter sector (i.e.\ the inflaton perturbations after restoring the four-dimensional diffeomorphsim invariance).
Assuming a canonical scalar field with a sound speed $c_s=1$, $S_{\rm matter}$
consistent with the  FLRW inflationary background is given by~\cite{Cheung-ml-2007st}
\begin{eqnarray}
\label{S-matter}
S_{\rm matter} &=& \int \! d^4 x  \sqrt{- g} \bigg[ -M^2_{P} \dot{H}(t+\pi) \bigg(\frac{1}{N^2} (1+\dot\pi-N^i\partial_i\pi)^2-
\partial^i\pi\partial_j\pi \bigg)  \nonumber\\
&&\qquad \qquad \quad - M^2_{ P} \left(3H^2(t+\pi) +\dot{H}(t+\pi)\right) \bigg]  \, ,
\end{eqnarray}
\looseness=-1
in which  $N$ and $N^i$ are the lapse and shift function in standard ADM formalism.   We note that the form of the above simple action is fixed by the background dynamics while all other  possible terms
are discarded in the gravitational decoupling limit. These discarded interactions include terms containing various combination of the extrinsic curvature $K_{ij}$ associated with the embedding of the three-surface $\phi(t)= constant$. As discussed in~\cite{Cheung-ml-2007st} the great advantage of EFT of inflation is in the decoupling limit where one can set $N=1$ and $N^i=0$ in the action  while neglecting the  perturbations from the $S_{\rm EH}$ action. In this limit the observational predictions such as the tilt of scalar perturbation  or the non-Gaussianity parameter $f_{NL}$ are determined to leading order  by the matter fluctuations. This was confirmed in non-attractor setup in~\cite{Akhshik-ml-2015nfa}. Finally, we comment that there can be other relevant interactions in action~(\ref{S-matter}) if one allows for a non-trivial sound speed $c_s \neq 1$, as studied in~\cite{Akhshik-ml-2015nfa}.

Our goal is to expand the EFT action to fourth order in perturbations of $\pi$. For this purpose,  we build upon the analysis of~\cite{Akhshik-ml-2015nfa} which calculated the action to cubic order in order to  calculate $f_{NL}$.

To obtain the EFT action  to fourth order, we have to expand the quantities $H(t + \pi)$ and $\dot H(t+\pi)$ to the corresponding orders as well.
To simplify the notation, let us define the order of differentiation via
$H^{(n)} \equiv \frac{d^n}{d t^n} H$. So for example, $\dot H= H^{(1)} = -\epsilon H^2$ and $\ddot H = H^{(2)} = -\dot{\epsilon}H^2-2\epsilon \dot{H}H$ and so on.
It is crucial to note that during the USR phase $\eta=-6$ so one can not neglect the higher derivatives of $H$. In addition, as shown in eq.~(\ref{eta-jump}),
immediately near the USR phase $\eta$ has a non-trivial behaviour so we should keep the derivative of $\eta$ in mind as well. On the other hand, as $\epsilon$ is very small, we only keep the first order contribution of $\epsilon$ in the interaction Hamiltonians. This is also consistent with our decoupling limit assumption where all gravitational back-reactions are discarded
to leading order of $\epsilon$.

With the above discussions in mind, the various derivatives of $H$ necessary to construct the action to fourth order of $\pi$ are given by
\begin{eqnarray}
{H^{(2)}}&=&-\dot{\epsilon}H^2-2\epsilon \dot{H}H \, , \nonumber \\
&=&-\eta \epsilon H^3 + 2\epsilon ^2 H^3
= -\eta \epsilon H^3 + O(\epsilon ^2),
\end{eqnarray}
where we have neglected the higher orders of $\epsilon$ in the last expression.
Continuing using this expansion, we have
\begin{eqnarray}
{H^{(3)}} &=& -\eta ^2 \epsilon H^4 - \epsilon  \frac{d \eta}{d t} H^3 + O(\epsilon ^2) \, ,\\
{H^{(4)}} &=& -\eta ^3 \epsilon H^5 - 3 \epsilon  \eta  \frac{d \eta}{d t} H^4
- \epsilon \frac{d^2 \eta}{d t^2} H^3
+O(\epsilon ^2) \, , \\
{H^{(5)}} &=& -\eta ^4 \epsilon H^6 - 6 \epsilon  \eta^2  \frac{d \eta}{d t} H^5
- 3 \epsilon  \left( \frac{d \eta}{d t}\right)^2 H^4 - 4 \epsilon  \eta  \frac{d^2 \eta}{d t^2} H^4
-   \epsilon    \frac{d^3 \eta}{d t^3} H^3
+O(\epsilon ^2) \, .
\end{eqnarray}
It is important to note that since there is a jump in $\eta$ from the USR phase to the third slow-roll limit, we can not neglect the derivates of $\eta$. As it is evident from eq.~(\ref{eta-jump}),  $\dot \eta$ contains a delta function which  will  contribute to the quartic interaction Hamiltonian as we shall see below.

With the above expansion at hands, we can calculate the quadratic, cubic and quartic actions. Starting with the quadratic action necessary to quantize the free theory we have
\begin{equation}
\label{S2}
S_2= M_P^2 \int dt d^3 x   a^3 \epsilon H^2 \big( \dot \pi^2 - (\partial_i \pi)^2 \big) \, ,
\end{equation}
The cubic action is given by
\begin{equation}
\label{S3}
S_{\pi^3} = - M_P^2 \int dt d^3 x   a^3 \bigg[ \bigg( H H^{(3)} + 3 \dot H  H^{(2)}  + \frac{1}{3} H^{(4)}\bigg) \pi^3  + H^{(3)} \pi^2 \dot \pi + H^{(2)} \pi \dot \pi^2  - H^{(2)} \pi \frac{(\partial \pi)^2}{a^2}
\bigg] \, ,
\end{equation}
while the quartic action is given by
\begin{eqnarray}
\label{S4}
S_{\pi^4}  =
 - M_P^2 \int dt d^3 x  a^3 &&\bigg[ \bigg( \dot H H^{(3)} + \frac{3}{4}   {\ddot H}^2  + \frac{1}{4} H H^{(4)}+ \frac{1}{12}  H^{(5)}  \bigg) \pi^4  + \frac{H^{(4)}}{3} \pi^3 \dot \pi  \nonumber\\
   &&~+ \frac{H^{(3)}}{2} \pi^2 \dot \pi^2 - H^{(3)} \pi^2 \frac{(\partial \pi)^2}{2a^2}
\bigg] \, .
\end{eqnarray}
We have included the effects of the gradient terms in cubic and quartic actions, as given by the last terms in big brackets in eqs.~(\ref{S3}) and~(\ref{S4}). As we shall see, the contributions of the interactions with the gradient terms are smaller compared to other terms on superhorizon scales, specially in the limit where $\Delta N , |h| \gg1$.

The above actions look problematic since in $H^{(5)}$ and $H^{(4)}$ we have $\ddot \eta$ and $\dddot \eta$ which contain the first and the second derivatives of
$\delta (\tau -\tau_e) $ respectively.
However, things become  simplified by noting that various terms in $S_{\pi^3}$ and $S_{\pi^4}$ group themselves  into total derivative terms which do not contribute into the Hamiltonian analysis.
More specifically, we have
\begin{equation}
\label{total1}
a^3 \bigg( H^{(3)} \pi^2 \dot \pi + \bigg( H H^{(3)} +  \frac{1}{3} H^{(4)}\bigg) \pi^3 \bigg) = \frac{1}{3} \frac{d}{d t}{\Big(  a^3 H^{(3)} \pi^3\Big)} \, .
\end{equation}
Discarding this total derivative and also neglecting the term containing $\dot H H^{(2)} \sim O(\epsilon^2)$ and going to conformal time the cubic action simplifies to
\begin{equation}
\label{action3}
S_{\pi^3} =  M_P^2 H^3 \int d\tau d^3 x\,  \eta \epsilon  a^2\,
\Big[  \pi \pi'^2  - \pi (\partial \pi)^2 \Big]  \, .
\end{equation}
Correspondingly, the cubic interaction Hamiltonian is obtained to be
%\begin{equation}
\begin{eqnarray}
\label{H3}
{\bf H}_3 &=& - M_P^2 H^3 \eta \epsilon a^2\, \int d^3 x  \Big[  \pi \pi'^2  - \pi (\partial \pi)^2 \Big]\, , \nonumber\\ 
&=&- M_P^2 H^3 \eta \epsilon a^2\, \int d^3 x  \Big[  \pi \pi'^2  +\frac{1}{2} \pi^2 \partial^2 \pi \Big] \, .
\end{eqnarray}
%\end{equation}
Here and below, a prime denotes the derivative with respect to the conformal
time. As expected, the above cubic Hamiltonian agrees exactly with those of~\cite{Akhshik-ml-2015nfa} when $c_s=1$.  Note that the  gradient term can not be ignored a priori, as its effects can be important~\cite{Akhshik-ml-2015nfa}.
 Note that the importance of the gradient interaction was highlighted in~\cite{Fumagalli-ml-2023hpa} as well.

Following a similar strategy for the quartic action, we note that
\begin{equation}
\label{total2}
a^3  \bigg( \frac{1}{4} H H^{(4)}+ \frac{1}{12}  H^{(5)}  \bigg) \pi^4  +  \frac{a^3 }{3} H^{(4)}\pi^3  \dot \pi  =  \frac{1}{12}  \frac{d}{dt}{\Big( a^3 H^{(4)} \pi^4 \Big)} \, .
\end{equation}
Neglecting the contribution of this total time derivative in the quartic action and
discarding the higher order term $\ddot H^2 \sim O(\epsilon^2)$, the quartic action to leading order of $\epsilon$ is obtained to be
\begin{equation}
\label{action4}
S_{\pi^4} =  \frac{M_P^2}{2} \int d\tau d^3 x\,   \epsilon
\bigg( \eta^2  a^2 H^4
+ a \frac{d \eta}{d \tau} H^3\bigg)\,    \Big[ \pi^2 \pi'^2 - \pi^2 (\partial \pi)^2 \Big]
\end{equation}
Note the important contribution from the term $\frac{d \eta}{d \tau} $ which causes a delta contribution in the interaction Hamiltonian as $\eta$ undergoes a jump from the USR phase to the third slow-roll phase as seen in eq.~(\ref{eta-jump}).

To calculate the quartic Hamiltonian, we have to be careful as the time derivative interaction $\pi' \pi^2$ in ${\bf H}_3$ induces a new term in the quartic Hamiltonian~\cite{Chen-ml-2006dfn, Chen-ml-2009bc}. More specifically, the quartic Hamiltonian receives a correction of the value $+M_P^2 H^4  \eta^2 \epsilon a^2\,    \pi^2 \pi'^2$ from the cubic action, so the total quartic Hamiltonian is obtained to be
\begin{equation}
\label{H4}
{\bf H}_4 =  \frac{M_P^2}{2}\int d^3 x \Big[
\Big( H^4 \eta^2 \epsilon a^2 - \eta'  H^3  \epsilon a \Big)  \pi^2 \pi'^2
+ \Big( H^4 \eta^2 \epsilon a^2 + \eta'  H^3  \epsilon a \Big)  \pi^2 (\partial \pi)^2
\Big]
\end{equation}
Equipped with the above interaction Hamiltonians, we can proceed to calculate the one-loop corrections in curvature perturbation power spectrum.

 However, note that we are interested in curvature perturbation on comoving surface $\calR$ while the above Hamiltonians are written in terms of the variable
 $\pi$. There are additional contributions from the non-linear relations between $\pi$ and $\calR$. To cubic order in $\pi$, the relation between $\calR$ and $\pi$ is given by~\cite{Jarnhus-ml-2007ia, Arroja-ml-2008ga}
 \begin{eqnarray}
 \label{pi-R1}
 \calR &=& - H \pi + \bigg( H \pi \dot \pi + \frac{\dot H}{2} \pi^2 \bigg)
 + \bigg( -H \pi \dot \pi^2 -\frac{H}{2} \ddot \pi \pi^2 - \dot H \dot \pi \pi^2 -\frac{\ddot H}{6} \pi^3 \bigg) \, , \nonumber\\
 &=& - H \pi + \frac{1}{2} \frac{d}{dt} \big(H \pi^2 \big) -  \frac{1}{6} \frac{d^2}{dt^2} \big(H \pi^3 \big) \, .
 \end{eqnarray}
\looseness=-1
 However, we calculate the two point correlation functions at the end of inflation
 $\tau=\tau_0 \rightarrow 0$ where it is assumed that the system is in the slow-roll regime and the perturbations are frozen on superhorizon scales, $\dot \pi =\ddot \pi=0$. Fortunately, in this limit all the non-linear corrections in $\calR$ in eq.~(\ref{pi-R1}) are suppressed  and we end up with the simple linear relation between $\calR$ and $\pi$:
 \begin{equation}
 \label{pi-R}
 \calR = - H \pi \, ,\quad  \quad \quad (\tau \rightarrow \tau_0).
 \end{equation}
Because of this linear relation between $\pi$ and $\calR$,  we can
 interchangeably use $\pi$ and $\calR$ in the following in-in integrals where the mode functions in the integrals are the solutions of the free theory (in the interaction picture).

 Also, from the second line of eq.~(\ref{pi-R1}) we see that the continuity of $\calR$ and $\dot \calR$ at the transition points  imply the continuity of $\pi$ and $\dot \pi$ as well. However, we note that $\ddot \calR$ may not be continuous across the transitions which also indicate that $\ddot \pi$ may not be continuous across the transition as well.

\section{The analysis  of bispectrum}
\label{bispectrum}

  As a prelude to calculate the one-loop corrections in power spectrum, in this  section we perform the bispectrum analysis to calculate $f_{NL}$ in the USR
  setup. This is a two-step model (USR $\rightarrow$ SR)  in which the first phase is USR inflation where the large CMB scales are generated followed by a second phase of slow-roll inflation. So this setup should not be confused with our main three-phases setup (SR $\rightarrow$ USR $\rightarrow$ SR)
  which was reviewed in section~\ref{setup} which will be employed for the loop correction in the next section. While the current analysis is new and interesting by itself, the reader who is only interested in one-loop corrections  can skip directly to the next section.

\looseness=-1
 The analysis of bispectrum for the simple USR $\rightarrow$ SR setup was performed originally in~\cite{Namjoo-ml-2012aa}, see also~\cite{Chen-ml-2013aj, Chen-ml-2013eea}.  This analysis was extended to a general non-attractor setup with $c_s \neq 1$ in~\cite{Akhshik-ml-2015nfa} using the EFT approach as in current analysis. However, the analysis in~\cite{Akhshik-ml-2015nfa} was for the case of extreme sharp transition $h \rightarrow -\infty$. Here, we extend the analysis of~\cite{Akhshik-ml-2015nfa} to the general case where the transition from USR phase to SR phase is controlled by the parameter $h \leq -6$. We confirm that our result for $f_{NL}$ is in agreement with the results of~\cite{Cai-ml-2018dkf} for a general sharp transition.
This can be viewed as a non-trivial confirmation of the validity of our EFT approach to calculate the one-loop corrections in power spectrum in the follow up section.

Going to Fourier space,  the comoving curvature perturbations $\calR$ is written as
\begin{equation}
\calR({\bf x}, t) = \int \frac{d^3 k}{(2\pi)^3} e^{i {\bf k}\cdot {\bf x}} \hat\calR_{\bf k}(t) \, ,
\end{equation}
 in which the operator $\hat\calR_{\bf k}(t)$ is written in terms of the creation and annihilation operators as $\hat\calR_{\bf k}(t)= \calR_k(t) a_{\bf k} + \calR^*_k(t) a_{-\bf k}^\dagger$. The creation and annihilation operators satisfy the usual commutation relations $[ a_{\bf k}, a^\dagger_{-\bf k'} ] = ( 2 \pi)^3 \delta (  {\bf k} + {\bf k'}) $.

The mode function  $\calR_k$ in the USR phase with the Bunch-Davies initial condition is given by~\cite{Namjoo-ml-2012aa}
\begin{equation}
\calR_{k} =  \frac{H}{ M_P\sqrt{4 \epsilon_i k^3}}  \bigg( \frac{\tau_i}{\tau} \bigg)^3
( 1+ i k \tau) e^{- i k \tau} \, ,
\end{equation}
in which $\epsilon_i$ is the value of the slow-roll parameter at the start of USR phase $\tau =\tau_i$. We note that on superhorizon scales where $k \tau \rightarrow 0$,  $\calR_k$ grows like  $\calR_k \propto \tau^{-3}$ which is the hallmark of the USR phase.  After the transition to the slow-roll  phase at $\tau =\tau_e$, the mode function during the final slow-roll phase ($\tau > \tau_e$) is given by
\begin{equation}
\label{R-alpha-beta}
\calR_{k} =  \frac{H}{ M_P\sqrt{4 \epsilon(\tau) k^3}} \Big[ \alpha_k ( 1+ i k \tau) e^{- i k \tau}  + \beta_k ( 1- i k \tau) e^{ i k \tau}  \Big]  \, ,
\end{equation}
with $\epsilon(\tau)$ given by eq.~(\ref{ep-N}) while
\begin{equation}
\label{alpha-beta}
\alpha_k = 1 + \frac{ i h}{ 4 k^3 \tau_e^3} ( 1 + k^2 \tau_e^2) \,,  \quad \quad
\beta_k= -\frac{i h}{ 4 k^3 \tau_e^3 } {( 1+ i k \tau_e)^2} e^{- 2 i k \tau_e} \, .
\end{equation}
It is instructive to look at the curvature perturbation power spectrum at the end of inflation $\tau_0 \rightarrow 0$. From the above expression for $\calR(\tau)$ we obtain
\begin{equation}
\label{PR-USR}
P_\calR(k, \tau_0) = \big| \calR_k(\tau_0)  \big|^2 = \frac{(h-6)^2}{ h^2}  \bigg(\frac{H^2}{4 k^3 M_P^2 \epsilon_e} \bigg) = \bigg(  1+ \frac{\epsilon(\tau_0)}{\epsilon(\tau_e)}
\bigg)^2  \bigg(\frac{H^2}{4 k^3M_P^2 \epsilon(\tau_0)} \bigg) \, .
\end{equation}
\looseness=-1
In the sharp transition with $h=-6$, we obtain $P_R(\tau_0)=  \Big(\frac{H^2}{ k^3M_P^2 \epsilon_e} \Big)= 4 P_\calR(\tau_e)$. Curiously, the power at the end of inflation is enhanced by a factor 4
 compared to the power at the end of USR phase. This is because the curvature perturbation is not frozen immediately after the USR phase and it keeps evolving till reaching the final attractor phase. On the other hand, for the extreme sharp transition with $h\rightarrow -\infty$ where the mode function freezes shortly after the USR phase the power at the end of inflation is approximately the same as the power at the end of USR phase as expected.  This difference in the behaviours of the mode function for $h=-6$ and $| h | \gg1$
 plays important roles in the final value of $f_{NL}$ measured at the end of inflation.

As the relation between $\pi$ and $\calR$ is linear when measured at the end of inflation  ($\calR = - H \pi)$, the three-point correlation calculated at the end of inflation is given by
\begin{eqnarray}
\big \langle \calR_{\bf p_1}(\tau_0)  \calR_{\bf p_2}(\tau_0)  \calR_{\bf p_3}(\tau_0)
\big \rangle &=& -H^3 \big \langle \pi_{\bf p_1}(\tau_0)  \pi_{\bf p_2}(\tau_0)  \pi_{\bf p_3}(\tau_0)     \big \rangle \nonumber\\
&=& 2 H^3   \mathrm{Im}  \int d \tau   \big \langle {\bf H}_3 (\tau) \pi_{\bf p_1}(\tau_0) \pi_{\bf p_2}(\tau_0)  \pi_{\bf p_3}(\tau_0)  \big\rangle  \, ,
\end{eqnarray}
yielding to
\begin{equation}
\big \langle \calR_{\bf p_1}  \calR_{\bf p_2}  \calR_{\bf p_3}
\rangle = - 4 M_P^2   \mathrm{Im}  \int d \tau \eta \epsilon a^2
 \big \langle
 \calR^*_{ p_1}(\tau_0) \calR^*_{ p_2}(\tau_0)  \calR^*_{ p_3}(\tau_0)
 \calR_{p_1}(\tau)  \calR'_{p_2}(\tau) \calR'_{p_3}(\tau) + 2 \mathrm{C. P.}
 \big \rangle  \end{equation}
in which $\mathrm{C. P.}$ stands for cyclic permutations. In addition, since we work on the superhorizon limit, we have neglected the gradient term in ${\bf H}_3$ in eq.~(\ref{H3}) to leading order of $q |\tau| \ll 1$.

 As the function $\eta(\tau)$ evolves from the  USR phase to the final slow-roll phase we divide the integral above into two regions $\tau_i < \tau < \tau_e$ with $\eta=-6$
 and $\tau_e < \tau < \tau_0$ with $\eta(\tau)$ given by eq.~(\ref{eta-N}). For the first contribution, we obtain
 \begin{equation}
 \label{first-region}
 \big \langle \calR_{\bf p_1}(\tau_0)  \calR_{\bf p_2}(\tau_0)  \calR_{\bf p_3}(\tau_0)
\big \rangle_{(1)} = - \frac{3(h-6)^2 (h+12)}{16 h^3}  \bigg(\frac{H^4}{M_P^4 \epsilon_e^2} \bigg) \bigg(\frac{1}{p_1^3 p_2^3} +  \frac{1}{p_1^3 p_3^3} + \frac{1}{p_2^3 p_3^3} \bigg) \, .
 \end{equation}
On the other hand, from the second contribution for the period  $\tau_e < \tau < \tau_0$  we obtain
 \begin{equation}
 \label{second-region}
 \big \langle \calR_{\bf p_1}(\tau_0)  \calR_{\bf p_2}(\tau_0)  \calR_{\bf p_3}(\tau_0)
\big \rangle_{(2)} =  \frac{3(h-6)^2 (h+6)}{8 h^3}  \bigg(\frac{H^4}{M_P^4 \epsilon_e^2} \bigg) \bigg(\frac{1}{p_1^3 p_2^3} +  \frac{1}{p_1^3 p_3^3} + \frac{1}{p_2^3 p_3^3} \bigg) \, .
 \end{equation}
The non-Gaussianity parameter $f_{NL}$ is defined via
\begin{equation}
 \big \langle \calR_{\bf p_1}(\tau_0)  \calR_{\bf p_2}(\tau_0)  \calR_{\bf p_3}(\tau_0)
\big \rangle \equiv  \frac{6}{5} f_{NL} \Big( P_\calR (p_1, \tau_0) P_\calR (p_2, \tau_0) +
2\,  \mathrm {cyclic \, \,   permutations }
\Big) \, .
\end{equation}
Using the value of the power spectrum   given by eq.~(\ref{PR-USR}) and combining the two contributions eqs.~(\ref{first-region}) and~(\ref{second-region}) we obtain
\begin{equation}
\label{fNL}
f_{NL} = \frac{5 h^2}{ 2 (h-6)^2} \, .
\end{equation}
The above result matches  exactly with the result obtained\footnote{As mentioned earlier, we work in the simplified limit where $\eta_V =0$ in the slow-roll phase while~\cite{Cai-ml-2018dkf} keeps this slow-roll parameter as well.} in~\cite{Cai-ml-2018dkf}.
In the limit of extreme sharp transition $h \rightarrow -\infty$, we obtain the original result in~\cite{Namjoo-ml-2012aa} with $f_{NL} =\frac{5}{2}$. On the other hand, for a sharp transition with $h=-6$ in which $\eta$ jumps to zero at the end of USR with  $\epsilon$ in the second phase held fixed to its value at the end of USR phase ($\epsilon(\tau_0) = \epsilon(\tau_e)$), we obtain $f_{NL} =\frac{5}{8}$.

A natural conclusion from the above analysis is that the value of $f_{NL}$ measured at the time of end of inflation depends crucially on the mechanism of transition. For the case of extreme sharp transition where the mode function freezes to its final slow-roll value immediately after the USR phase, one can use the value of $f_{NL}$ calculated right before or right after the USR phase as employed in~\cite{Namjoo-ml-2012aa} and~\cite{Akhshik-ml-2015nfa}. However, if the mode function keeps evolving after the USR phase (including the ``natural'' case with $h=-6$), it is crucial to make the measurement at the end of inflation. As demonstrated in~\cite{Cai-ml-2018dkf}, the maximum value of $f_{NL}$ obtained in this setup is
$f_{NL}= \frac{5}{2}$.

From the above exercise  we conclude that our EFT approach is trusted to calculated the loop corrections
in power spectrum since we are interested in leading corrections from the modes which are growing in the USR phase while discarding the subleading slow-roll corrections.
This is the limit where the EFT formalism is guaranteed to be trusted.

\section{One-loop correction in power spectrum}
\label{one-loop}

 After the  warmup exercise to calculate the bispectrum in USR $\rightarrow$ SR setup using our EFT formalism, we are now in a position to calculate the one-loop corrections in power spectrum in our SR $\rightarrow$ USR $\rightarrow$ SR setup.
 As mentioned earlier, for a consistent treatment we have to include the corrections from both the cubic Hamiltonian ${\bf H}_3$ and the quartic Hamiltonian ${\bf H}_4$. Somehow, it was concluded in~\cite{Kristiano-ml-2022maq} that there is no contribution from the quartic self interaction of curvature perturbation and the quartic action is suppressed by higher powers of $\epsilon$~\cite{Jarnhus-ml-2007ia}. While there are non-linear relations between $\calR$ (or notation $\zeta$ used in~\cite{Kristiano-ml-2022maq} ) but our
 specific calculation shows that ${\bf H}_4$ is actually non-zero and has the same order of $\epsilon$ as ${\bf H}_3$. Looking at the structure of ${\bf H}_4$
 given in eq.~(\ref{H4}) we see that it contains two terms. The first term make contribution in the ``bulk'' either in the region $\tau_i <\tau< \tau_e$ or $\tau_e < \tau<  \tau_0$ while the second term is a localized delta term $\delta (\tau-\tau_e)$
 which makes a contribution right at the transition time $\tau_e$ as given in eq.~(\ref{eta-jump}).

 We present the analysis for a general value of $h$ for the sharp transitions. However, the analysis of~\cite{Kristiano-ml-2022maq} is for the case $h=-6$ when
 the value of the slow-roll parameter in the third slow-roll phase  is fixed to its  value at the end of USR phase,  $\epsilon(\tau_0) = \epsilon(\tau_e)$.

 To calculate the loop corrections, we employ the standard in-in formalism~\cite{Weinberg-ml-2005vy} in which the expectation value of the operator $\hat {O}[\pi]$ at the end of inflation $\tau_0$ is given by the perturbative Dyson series,
 \begin{equation}
 \label{Dyson}
 \langle \hat O[\pi(\tau_0)] \rangle = \bigg \langle \bigg[ \bar {\mathrm{T}} \exp \bigg( i \int_{-\infty}^{\tau_0} d \tau' H_{in} (\tau') \bigg) \bigg] \,  \hat O[\pi(\tau_0)]  \, \bigg[ \mathrm{T} \exp \bigg( -i \int_{-\infty}^{\tau_0} d \tau' H_{in} (\tau') \bigg) \bigg]
 \bigg \rangle
 \end{equation}
 in which $\mathrm{T}$ and $\bar {\mathrm{T}}$ represents the time ordering and anti-time ordering respectively while $H_{in}(t)$ represents the interaction Hamiltonian
 which in our case is $H_{in}(\tau) = {\bf H}_3 + {\bf H}_4$.

 To calculate the in-in analysis we need the mode function for $\pi$ (or $\calR)$ during the USR and the follow up  slow-roll phase.
 Imposing the Bunch-Davies initial condition during the first slow-roll stage
 (i.e.\ for $\tau < \tau_i$) we have
 \begin{equation}
\calR^{(1)}_{k} =  \frac{H}{ M_P\sqrt{4 \epsilon_i k^3}}
( 1+ i k \tau) e^{- i k \tau} \, , \quad \quad (\tau < \tau_i) \, .
\end{equation}
in which $\epsilon_i$ is the value of the slow-roll parameter at the start of inflation
when the CMB scale mode has left the horizon. The superscript (1) indicate the first phase. During the USR phase, the mode function is given formally by an expansion similar to eq.~(\ref{R-alpha-beta}):
\begin{equation}
\calR^{(2)}_{k} =  \frac{H}{ M_P\sqrt{4 \epsilon_i k^3}}  \bigg( \frac{\tau_i}{\tau} \bigg)^3
\Big[ \alpha^{(2)}_k ( 1+ i k \tau) e^{- i k \tau}  + \beta^{(2)}_k ( 1- i k \tau) e^{ i k \tau}  \Big]  \, ,
\end{equation}
with the coefficients $\alpha^{(2)}_k$ and $\beta^{(2)}_k$ given by
\begin{equation}
\label{alpha-beta2}
\alpha^{(2)}_k = 1 + \frac{3 i }{ 2 k^3 \tau_i^3} ( 1 + k^2 \tau_i^2) \, \quad \quad
\beta^{(2)}_k= -\frac{3i }{ 2 k^3 \tau_i^3 } {( 1+ i k \tau_i)^2} e^{- 2 i k \tau_i} \, .
\end{equation}
Finally, imposing the continuity of the mode function and its time derivative
at $\tau_e$, the mode function in the final slow-roll phase, denoted by the superscript (3), is obtained to be
\begin{equation}
\calR^{(3)}_{k} =  \frac{H}{ M_P\sqrt{4 \epsilon(\tau) k^3}}
\Big[ \alpha^{(3)}_k ( 1+ i k \tau) e^{- i k \tau}  + \beta^{(3)}_k ( 1- i k \tau) e^{ i k \tau}  \Big] \, ,
\end{equation}
with $\epsilon(\tau)$ given by eq.~(\ref{ep-N}) while
the coefficients $\alpha^{(3)}_k$ and $\beta^{(3)}_k$ are given by,
$$
\label{alpha-beta3}
\alpha^{(3)}_k = \frac{1}{8 k^6 \tau_i^3 \tau_e^3}  \Big[ 3h
 ( 1 -i k \tau_e)^2 (1+i k \tau_i)^2 e^{2i k (\tau_e- \tau_i)}
-i (2 k^3 \tau_i^3 + 3i k^2 \tau_i^2 + 3 i) (4 i k^3 \tau_e^3- h k^2 \tau_e^2 - h) \Big]
\nonumber
$$
and
{\rdddmathspace$$
\beta^{(3)}_k=   \frac{-1}{8 k^6 \tau_i^3 \tau_e^3}  \Big[ 3 ( 1+ i k \tau_i)^2 ( h+ h k^2 \tau_e^2 + 4 i k^3 \tau_e^3 ) e^{-2 i k \tau_i} + i h ( 1+ i k \tau_e)^2  ( 3 i + 3 i k^2 \tau_i^2 + 2 k^3 \tau_i^3 ) e^{- 2 i k \tau_e}
 \Big] \nonumber
$$}\relax

Armed with the above mode functions, we are ready to calculate the one-loop corrections in power spectrum. To clarify the notation, the momentum for the
 CMB modes are denoted by  ${\bf p}_1$ and ${\bf p}_2$ while  the momentum
 for the small scale modes running in the loop are denoted by ${\bf q}$.
Since we are interested in only to the modes which leave the horizon during the USR phase, we cut the loop integrals in the intervals $ q_i \leq q < q_e $ in which $q_i = -\frac{1}{\tau_i}$ and $q_e= - \frac{1}{\tau_e}$. The number of e-folds of the USR phase  $\Delta N= N(\tau_e) - N(\tau_i)$ is related to $q_i$ and $q_e$ via
\begin{equation}
\label{delta-N-def}
e^{- \Delta N}=  \frac{\tau_e}{\tau_i} = \frac{q_i}{q_e}  \, .
\end{equation}
For a rapid growth of the mode function during USR phase  we typically require    $\Delta N >1$.

\subsection{Corrections from quartic self interaction}
\label{quartic-H}

 We start with the loop corrections from the quartic self interaction eq.~(\ref{H4}) as
 the in-in integrals are easier to handle.

 The one-loop correction for the CMB-scale mode ${\bf p}_1$ and $\bfp_2$ induced from
 $\Hb$ is given by
 \begin{equation}
 \label{H4-in-in}
 \langle \calR_{\bfp_{1}}(\tau_0) \calR_{\bfp_{2}}(\tau_0) \rangle
 = H^2  \langle \pi_{\bfp_{1}}(\tau_0) \pi_{\bfp_{2}}(\tau_0) \rangle
 = - 2 H^2   \mathrm{Im}  \int_{-\infty}^{\tau_0} d \tau   \big \langle {\bf H}_4 (\tau) \pi_{\bf p_1}(\tau_0) \pi_{\bf p_2}(\tau_0)  \big\rangle \, .
 \end{equation}

As the ${\bf H}_4$  term has contributions from the bulk $ \tau_i<  \tau<  \tau_0$ and from the source term $\delta (\tau-\tau_e)$, we have two different types of in-in integrals: in-in performed in the bulk and in-in performed at $\tau=\tau_e$. Let us first start with the bulk contribution.

After performing all the contractions, we end up with integral over the momentum
$\bfq$ running in the loop yielding to three different contributions, which we denote by terms $A, B$  and $\tilde A$ as follows
 \begin{equation}
 \langle \calR_{\bfp_{1}}(\tau_0) \calR_{\bfp_{2}}(\tau_0) \rangle|_{\mathrm{bulk}}
 \equiv   (2 \pi)^3 \delta^3 (\bfp_1 + \bfp_2)  (A + B+ \tilde A) \, ,
 \end{equation}
 in which  $A$ and $B$ are calculated from the interaction $\pi^2 \pi'^2$ while $\tilde A$
 is calculated from the gradient interaction $\pi^2 (\partial \pi)^2$ in ${\bf H_4}$.
   More specifically, we have
 \begin{equation}
 \label{A-term}
 A= - 2 M_P^2  \int \frac{d^3 \bfq}{(2 \pi)^3}
 \int  d \tau \left( \eta^2 - \frac{\eta'}{a H}  \right) \epsilon(\tau) a(\tau)^2  | \calR'_q(\tau)|^2
 \mathrm{Im} \Big[  \calR_\bfp^*(\tau_0)^2  \calR_\bfp(\tau)^2    \Big] \, ,
 \end{equation}
 and
\begin{equation}
 \label{B-terma}
 B= - 8 M_P^2  \int \frac{d^3 \bfq}{(2 \pi)^3}
 \int  d \tau \left( \eta^2 - \frac{\eta'}{a H}  \right) \epsilon(\tau) a(\tau)^2
 \mathrm{Im} \Big[  \calR_\bfp^*(\tau_0)^2  \calR_\bfp(\tau) \calR'_\bfp(\tau) \calR_\bfq(\tau) \calR'_\bfq(\tau)^*     \Big] \, ,
 \end{equation}
 while
 \begin{equation}
 \label{tildeA-term}
 \tilde A= - 2 M_P^2  \int \frac{d^3 \bfq}{(2 \pi)^3}
 \int  d \tau \left( \eta^2 + \frac{\eta'}{a H}  \right) \epsilon(\tau) a(\tau)^2 q^2 | \calR_q(\tau)|^2
 \mathrm{Im} \Big[  \calR_\bfp^*(\tau_0)^2  \calR_\bfp(\tau)^2    \Big] \, .
 \end{equation}
 We comment that we need to insert the mode function of $\pi$ in eq.~(\ref{H4-in-in})
 as our quartic action is in terms of $\pi$ and not $\calR$. However, since both $\calR$ and $\pi$ share the same mode function at the level of free theory (note that $\calR = - H \pi + {\cal O} (\pi^2) )$, then we have inserted the mode function of $\calR$ in the above expressions for $A, B$ and $\tilde A$, with the understanding that  $\calR = - H \pi$ as in free theory.

 There are two different contributions in the bulk: during the USR phase with  $\tau_i < \tau < \tau_e$ and after the USR phase $\tau_e < \tau < \tau_0$.
 During the first interval, $\eta =-6$ and $\eta' =0$ and we calculate the corresponding in-in integrals in details below.
On the other hand,  during the second bulk phase, we approximately have
 \begin{equation}
   \eta^2 \mp \frac{\eta'}{a H} \simeq \pm \frac{18 (6+h) \tau^3}{h \tau_e^3} + O(\tau^5)\, .
 \end{equation}
 Therefore, we see that this combination falls off rapidly after the USR phase.
Furthermore, as the system reaches the attractor phase rapidly after the transition, the mode function $\calR_\bfq(t)$ approaches its usual slow-roll
form and we will have additional  suppressions from the time derivatives of the mode function. Correspondingly, one expects the contribution from the final slow-roll region
 $\tau_e < \tau < \tau_0$ to be much smaller than those coming from the USR phase. We have checked explicitly that this is indeed the case. Therefore, in the rest of the analysis we consider the contribution from the first bulk region during the USR phase.

 During the USR phase, the superhorizon modes grow like $\tau^{-3}$ and
 \begin{equation}
 \label{Rprime-R}
 |\calR'_q(\tau)|^2 \simeq  \frac{9}{\tau^2} |\calR_q(\tau)|^2 \simeq
 \frac{9}{\tau^2} \bigg(\frac{\tau_e}{\tau} \bigg)^6  |\calR_q(\tau_e)|^2 \, .
 \end{equation}
On the other hand,
\begin{equation}
\label{Im1}
\mathrm{Im} \Big(  \calR_\bfp^*(\tau_0)^2  \calR_\bfp(\tau)^2 )   \Big)
=\frac{1}{24} \frac{H^4}{M_P^4 \epsilon_i^2 p^3} \frac{\tau_i^6}{ \tau_e^3 \tau^3}
\bigg( \frac{6-h}{h} \tau^3 + \tau_e^3 \bigg) \, .
\end{equation}
Plugging the expressions~(\ref{Rprime-R}) and~(\ref{Im1}) in eq.~(\ref{A-term}) with
$\eta=-6$ we end up with a time integral of the following form for the A-term:
\begin{equation}
\int_{\tau_i}^{\tau_e} d \tau \bigg(  \frac{6-h}{h} \tau^{-4} + \tau_e^3 \tau^{-7}
\bigg) = \bigg( \frac{1}{6} -\frac{2}{h} \bigg) \tau_e^{-3} \, .
\end{equation}
Plugging all numerical factors together, we finally obtain
\begin{equation}
A= 27 \bigg(  \frac{2}{h}- \frac{1}{6} \bigg)  \frac{H^2}{M_P^2 \epsilon_i p^3}
\int \frac{d^3 \bfq}{(2 \pi)^3}   |\calR_q(\tau_e)|^2 \, .
\end{equation}
To perform the above loop integral we restrict ourselves to modes which become
superhorizon during the USR phase with $q_i < q <q_e$. Also,
the power spectrum at the end of USR phase is given by $P_\calR(\tau_e) = |\calR_q(\tau_e)|^2 =   \frac{H^2  }{4 M_P^2 \epsilon_i q^3}e^{6 \Delta N} $.
Performing the remaining loop integral over $q$ yields the factor $\frac{1}{2 \pi^2} \ln(\frac{q_e}{q_i}) = \frac{\Delta N}{2 \pi^2} $. Combining all results together, we have
\begin{equation}
\label{A-val}
 A= \bigg( \frac{27}{2h} -\frac{9}{8}  \bigg) \big( {\Delta N}{e^{6 \Delta N}} \big)  \frac{H^4  }{ 2 \pi^2 M_P^4 \epsilon_i^2 p^3 }  \, .
\end{equation}

To calculate the B-term, we note that to leading order in $p$,
\begin{equation}
\label{Im2}
 \calR_\bfp^*(\tau_0)^2  \calR_\bfp(\tau) \calR'_\bfp(\tau)  = - \frac{i  H^4}{16 M_P^4 \epsilon_i^2 p^3} \, \frac{\tau_i^6 }{ \tau^4} \, .
\end{equation}
As the above combination is imaginary by itself, the remaining factor
in the B-term simplifies to
\begin{equation}
\calR_\bfq(\tau) \calR'_\bfq(\tau)^*  = -\frac{3}{\tau} |\calR_q(\tau)|^2
=  -\frac{3}{\tau}  \bigg(\frac{\tau_e}{\tau} \bigg)^6  |\calR_q(\tau_e)|^2 \, .
\end{equation}
 Curiously, from the above expressions we realize that there is no explicit dependence on the parameter $h$ in the B-term. Performing a simple time integral and the momentum integral over $q$ as in the case of A-term, we finally obtain
 \begin{equation}
 \label{B-val}
 B= \frac{9}{4}   \big( {\Delta N}{e^{6 \Delta N}} \big) \frac{H^4  }{
 2 \pi^2 M_P^4 \epsilon_i^2 p^3 }  \, .
 \end{equation}

Finally, the last contribution in the bulk is $\tilde A$, which comes from
the interaction containing the gradient term. As we calculate the loop effects from the superhorizon modes, one expects that the contributions from the gradient term to be subleading. More specifically, on superhorizon scales $(\partial \calR)^2 \simeq q^2 \calR^2$  while $\calR'^2 \simeq \frac{9}{\tau^2} \calR^2$. Correspondingly the interaction term $\calR^2 (\partial \calR)^2 $ is suppressed  compared to the time derivative interaction $\calR^2 \calR'^2$ by a factor $\frac{q^2 \tau^2}{9}$.  With this discussion in mind, let us calculate $\tilde A$ given in eq.~(\ref{tildeA-term}). Fortunately, the integral in $\tilde A$ is very similar to the integral
in $A$, yielding
\begin{equation}
\label{A-tilde-val}
\tilde A = \frac{3(12-h)}{32 h} {e^{6 \Delta N}}  \frac{H^4  }{
 2 \pi^2 M_P^4 \epsilon_i^2 p^3 }  \, .
\end{equation}
\looseness=-1
Curiously, we see that there is no factor $\Delta N$ in $\tilde A$ compared
to the $A$ and $B$ terms. This is in line with the above discussions that one typically expects that $\tilde A$ to be subleading compared to $A$ and $B$ terms
specially in the limit when $\Delta N \gg 1$. However, for a limited duration of USR phase with $\Delta N \lesssim 1$, one may not neglect the contributions of $\tilde A$ compared to other terms.

\looseness=-1
After calculating the bulk contributions, now we calculate the localized terms
 induced by $\eta'$ at $\tau= \tau_e$. Again, we have three contributions from the source terms denoted by terms $C, D$ and $\tilde C$. The term $C$ is similar to term $A$, the term $D$ is similar to term $B$ while $\tilde C$ is similar to $\tilde A$.
It is understood that  these terms occur because $\eta' = - h \delta (\tau -\tau_e)$. More specifically,
  \begin{align}
 \label{C-term}
 C&= - 2 h M_P^2  \int \frac{d^3 \bfq}{(2 \pi)^3}
  \epsilon(\tau_e) a(\tau_e)  | \calR'_q(\tau_e)|^2
 \mathrm{Im} \Big[  \calR_\bfp^*(0)^2  \calR_\bfp(\tau_e)^2    \Big] \, ,
 \\
 \label{B-term}
 D&= - 8 h M_P^2  \int \frac{d^3 \bfq}{(2 \pi)^3}
 \epsilon(\tau_e) a(\tau_e)
 \mathrm{Im} \Big[  \calR_\bfp^*(\tau_0)^2  \calR_\bfp(\tau_e) \calR'_\bfp(\tau_e) \calR_\bfq(\tau_e) \calR'_\bfq(\tau_e)^*     \Big] \, .
 \end{align}
 and
 \begin{equation}
 \label{tildeC-term}
 \tilde C=  2 h M_P^2  \int \frac{d^3 \bfq}{(2 \pi)^3}  q^2
  \epsilon(\tau_e) a(\tau_e)  | \calR_q(\tau_e)|^2
 \mathrm{Im} \Big[  \calR_\bfp^*(0)^2  \calR_\bfp(\tau_e)^2    \Big] \, .
 \end{equation}

Performing the loop integrals as in the bulk cases and, using eqs.~(\ref{Im1}) and~(\ref{Im2}) now calculated at $\tau= \tau_e$, we obtain
\begin{equation}
\label{C-val}
  C=  \frac{9}{8} \big(  {\Delta N}{e^{6 \Delta N}} \big) \frac{H^4  }{ 2 \pi^2 M_P^4 \epsilon_i^2 p^3 } \, ,
\end{equation}
 and
 \begin{equation}
\label{D-val}
 D=  \frac{3 h}{8}  \big( {\Delta N}{e^{6 \Delta N}} \big) \frac{H^4  }{ 2 \pi^2 M_P^4 \epsilon_i^2 p^3 } \, .
\end{equation}
Curiously, we see that  the $D$ term has an explicit  factor of $h$ while this
is not the case for the $C$ term.

On the other hand, for the
 $\tilde C$ we obtain
 \begin{equation}
 \label{tildeC-val}
  \tilde C = -\frac{3}{32 } {e^{6 \Delta N}}  \frac{H^4  }{
 2 \pi^2 M_P^4 \epsilon_i^2 p^3 }  \, .
 \end{equation}
As expected, the $\tilde C$ term is somewhat suppressed compared to $C$ and $D$ terms specially in the limit when $\Delta N \gg 1$.

Summing the leading terms  $A+ B+ C+D$, the total contribution from the quartic Hamiltonian corresponding to the right panel of figure~\ref{Feynman-fig}
 is obtained to be
\begin{equation}
\label{quartic-power}
  \langle \calR_{\bfp_{1}}(\tau_0) \calR_{\bfp_{2}}(\tau_0) \rangle_{\Hb}
 \simeq  \frac{3}{8h} \big(  h^2 + 6h + 36
 \big) \big( {\Delta N}{e^{6 \Delta N}}  \big) \frac{H^4  }{ 2 \pi^2 M_P^4 \epsilon_i^2 p^3 } \,
 ( 2 \pi)^3 \delta^3 (\bfp_1 + \bfp_2) \, .
\end{equation}
 We notice that the one-loop correction is a non-linear function of the parameter $h$
 and grows linearly for $| h| \gg 1$.

 On the other hand, the contribution of the subleading terms $\tilde A$ and $\tilde C$ are given by
 \begin{equation}
\label{quartic-power2}
  \langle \calR_{\bfp_{1}}(\tau_0) \calR_{\bfp_{2}}(\tau_0) \rangle_{\Hb}^{\mathrm{subleading}}
 \simeq   \frac{6-h}{16 h}
  {e^{6 \Delta N}}  \frac{H^4  }{ 2 \pi^2 M_P^4 \epsilon_i^2 p^3 } \,
 ( 2 \pi)^3 \delta^3 (\bfp_1 + \bfp_2) \, .
\end{equation}
 As this contribution is subleading in both limits $|h|, \Delta N \gg 1$ compared to eq.~(\ref{quartic-power}), we do not consider its effects in our final estimation below.

\subsection{Corrections from cubic self interaction}
\label{cubic-H}

 Now we calculate the loop correction from the cubic interaction $\Ha$ corresponding to the left panel of figure~\ref{Feynman-fig}. There are noticeable differences compared to the quartic case. First, one simplification is that there is no source term as there is no localized term from $\eta'$ in $\Ha$. However, a major complication is that the integrals in the bulk become nested (i.e.\ double integrals).
We present the detail technicalities of the nested integrals in appendix~\ref{appendix} and here outline the main steps leading to the final result.

Expanding the Dyson series to second order in $\Ha$ we have
\begin{equation}
   \langle \calR_{\bfp_{1}}(\tau_0) \calR_{\bfp_{2}}(\tau_0) \rangle_{\Ha} =    \langle \calR_{\bfp_{1}}(\tau_0) \calR_{\bfp_{2}}(\tau_0) \rangle_{(2,0)} +   \langle \calR_{\bfp_{1}}(\tau_0) \calR_{\bfp_{2}}(\tau_0) \rangle_{(1,1)} +    \langle \calR_{\bfp_{1}}(\tau_0) \calR_{\bfp_{2}}(\tau_0) \rangle_{(0, 2)}
  \end{equation}
  in which
\begin{eqnarray}
  \label{20-int}
\langle \calR_{\bfp_{1}}(\tau_0) \calR_{\bfp_{2}}(\tau_0) \rangle_{(2,0)} &=&
- \int_{-\infty}^{\tau_0} d \tau_1 \int_{-\infty}^{\tau_1} d \tau_2
\big \langle \Ha (\tau_2)  \Ha (\tau_1) \calR_{\bfp_{1}}(\tau_0) \calR_{\bfp_{2}}(\tau_0)
 \big \rangle   \nonumber\\
&=& \langle \calR_{\bfp_{1}}(\tau_0) \calR_{\bfp_{2}}(\tau_0)
\rangle^\dagger_{(0,2)}\, ,
\end{eqnarray}
and
 \begin{equation}
  \label{11-int}
\langle \calR_{\bfp_{1}}(\tau_0) \calR_{\bfp_{2}}(\tau_0) \rangle_{(1,1)} =
 \int_{-\infty}^{\tau_0} d \tau_1 \int_{-\infty}^{\tau_0} d \tau_2
\big \langle \Ha (\tau_1)   \calR_{\bfp_{1}}(\tau_0) \calR_{\bfp_{2}}(\tau_0)
\Ha (\tau_2) \big \rangle   \, .
\end{equation}

As in the case of quartic self interaction, the main contributions in the in-in integrals come from the USR phase and the contributions from the final slow-roll phase are negligible. As explained before, this is because  the mode functions approach their slow-roll values and their derivatives falls off exponentially during the slow-roll phase.

Combining all contributions (see appendix~\ref{appendix} for further details) we obtain
\begin{equation}
\label{cubic-rsult}
   \langle \calR_{\bfp_{1}}(\tau_0) \calR_{\bfp_{2}}(\tau_0)  \rangle_{\Ha} =
     - 8 M_P^4  \int \frac{d^3 \bfq}{(2 \pi)^3}
     \int_{\tau_i}^{\tau_e} d \tau_1
    \int_{\tau_i}^{\tau_1} d \tau_2 \,  {\cal F} (\tau_1, \tau_2; q)
         \end{equation}
 in which the integrand ${\cal F} (\tau_1, \tau_2; q)$ is given by
 \begin{equation}
  {\cal F} (\tau_1, \tau_2; q) \equiv  \mathrm{Im} \left\{ X^*_1(  \tau_2) \Big[  \Big( 1- f^*_q(\tau_2)  \Big) \Big( \, \delta (\tau_1) + 2 \beta ( \tau_1)\,  \Big)     
  - f_q(\tau_1) \delta (\tau_1)  \Big] \right\}
 \end{equation}
 where
 \begin{eqnarray}
 \delta (\tau) &\equiv& 2 \epsilon \eta a^2 \calR'_q(\tau)^2 \mathrm{Im}
 \big[  \calR_p^*(\tau_0)  \calR_p(\tau)   \big] \, , \nonumber\\
 \beta (\tau) &\equiv& 2 \epsilon \eta a^2 \calR'_q(\tau) \calR_q(\tau)
 \mathrm{Im}
 \big[  \calR_p^*(\tau_0)  \calR'_p(\tau)   \big] \, ,
 \end{eqnarray}
 and
 \begin{equation}
 X_1(\tau) \equiv  \epsilon \eta a^2  \calR_p^*(\tau_0)  \calR_p(\tau) \calR'_q(\tau)^2 \, .
 \end{equation}
In addition, the function $f_q(\tau_i)$ represents the contribution from the gradient term which is defined as
\begin{equation}
f_q(\tau)  \equiv \frac{\big( \partial \calR_q \big)^2}{  \calR_q'^2} = \frac{q^2 \calR_q^2}{\calR_q'^2}
\simeq \frac{(q\tau)^2}{9} \, ,
\end{equation}
\looseness=-1
where the latter approximation is obtained in the superhorizon limit.
From the above expression we see that on the superhorizon scales  $f_q(\tau) <1$. However, since we are dealing with a nested integral which may be accumulative,
one can not  conclude a priori that the contributions of the gradient interactions are subleading (unlike in the case of quartic interaction).

Performing the in-in nested integrals is more challenging compared to the quartic case. The lower bound on the $\tau_2$ and $\tau_1$ integrations  in Eq. (\ref{cubic-rsult}) is $\tau_i$. However, we choose the strategy to integrate only over the modes which become superhorizon during the USR phase. For a given momentum $q$ running in the loop, this corresponds to integrating over  $-\frac{1}{q} \leq \tau_2 \leq \tau_1\leq \tau_e$. 
Calculating the above integrals (see appendix A for details)   the final result to leading order is obtained to be
\begin{equation}
\label{cubic-power}
    \langle \calR_{\bfp_{1}}(\tau_0) \calR_{\bfp_{2}}(\tau_0)  \rangle_{\Ha} =
    \frac{ 27 ( h + 8)}{4h}
   \big( \Delta N e^{6 \Delta N} \big)
 \frac{H^4  }{  2 \pi^2 M_P^4 \epsilon_i^2 p^3 }
  ( 2 \pi)^3 \delta^3 (\bfp_1 + \bfp_2)  \, .
\end{equation}

We comment that if we do not impose the requirement that the mode to be superhorizon then the lower bound of integration in Eq. (\ref{cubic-rsult}) is
$\tau_i$. Then,   independent of the value of the  mode $q$, we should integrate over the range $\tau_i \leq \tau_2 \leq \tau_1\leq \tau_e$. Performing the nested integral using this prescription modifies the prefactor 
in Eq. (\ref{cubic-power}) such that 
%$ \frac{ 27 ( h + 8)}{4h} \rightarrow  \frac{9 ( h -12)}{8h}$. While this changes the numerical factors in the final loop correction but it does not change the result qualitatively. 
\begin{equation}
\label{cubic-power-b}
    \langle \calR_{\bfp_{1}}(\tau_0) \calR_{\bfp_{2}}(\tau_0)  \rangle_{\Ha} =
    \frac{9 ( h -12)}{8h}
   \big( \Delta N e^{6 \Delta N} \big)
 \frac{H^4  }{  2 \pi^2 M_P^4 \epsilon_i^2 p^3 }
  ( 2 \pi)^3 \delta^3 (\bfp_1 + \bfp_2)    \quad 
 \mathrm{(second \, \, strategy)}  .
\end{equation}
Having said this, we believe that the first strategy is more physical since only the modes which become superhorizon during the USR phase are expected to affect the long mode. However, this does  not change the result significantly  as  both Eqs. (\ref{cubic-power}) and (\ref{cubic-power-b}) qualitatively look very similar.

As in the quartic case, we see a non-linear dependence on the parameter $h$. However, the contribution from the cubic Hamiltonian becomes negligible compared to the quartic Hamiltonian  in the limit of $| h| \gg 1$.

\subsection{Total one-loop corrections }
\label{Total}

\sloppy{Having calculated the one-loop corrections from the quartic and cubic Hamiltonians, eqs.~(\ref{quartic-power}) and~(\ref{cubic-power}), the total one-loop corrections
in power spectrum is given by (after neglecting the trivial factor
$( 2 \pi)^3 \delta^3 (\bfp_1 + \bfp_2)$)}
\begin{equation}
\label{loop-power}
\Delta \calP(\bfp) \equiv    \frac{p^3}{2 \pi^2}   \langle \calR_{\bfp_{1}}(\tau_0) \calR_{\bfp_{2}}(\tau_0) \rangle_{\mathrm{1-loop}} =\frac{6}{ h} \big(  h^2 + 24 h + 180) (\Delta N e^{6 \Delta N}) \calP_{\mathrm{CMB}}^2 \, ,
\end{equation}
in which $\calP_{\mathrm{CMB}} =  \frac{H^2  }{ 8 \pi^2 M_P^2 \epsilon_i } \sim  2\times 10^{-9}$ is the power spectrum on the CMB scale.

Eq.~(\ref{loop-power}) is the main result of this paper. As mentioned before, we see a nonlinear dependence of the one-loop correction to the parameter $h$. Note that
this result obtained in the limit of sharp transition so eq.~(\ref{loop-power}) is not applicable for the limit  $|h| \ll  1$.

In passing, we comment that if we follow the second strategy and integrate over all modes with $\tau_i \leq \tau_2 \leq \tau_1\leq \tau_e$ (i.e. the lower bound of the nested time integral is fixed  to be $\tau_i$), then combining eqs.~(\ref{quartic-power}) and~(\ref{cubic-power-b}) yields  
\begin{equation}
\label{loop-power-b}
\Delta \calP(\bfp)  = (6 h + 54)  (\Delta N e^{6 \Delta N}) \calP_{\mathrm{CMB}}^2   \quad \quad 
 \mathrm{(second \, \, strategy)}  .
\end{equation}
However, as mentioned previously, we follow the first strategy while 
both results in eqs.~(\ref{loop-power}) and eq.~(\ref{loop-power-b}) are 
physically similar, except in numerical prefactor.  

\looseness=-1
For the particular case $h=-6$, where $\epsilon$ in the final slow-roll stage keeps its value at the end of USR phase, from eq.~(\ref{loop-power}) we obtain
$\Delta \calP = -72 (\Delta N e^{6 \Delta N}) \calP_{\mathrm{CMB}}^2$. This is the limit which we can compare our result with the results of~\cite{Kristiano-ml-2022maq}. In their notation,  $\Delta \eta(\tau_e) =-h=6$ and up to a model dependent  constant they would obtain  $\Delta \calP = \frac{\Delta \eta(\tau_e)^2}{4} (\Delta N e^{6 \Delta N}) \calP_{\mathrm{CMB}}^2 = 9 (\Delta N e^{6 \Delta N}) \calP_{\mathrm{CMB}}^2$. Therefore, our result is larger than that of~\cite{Kristiano-ml-2022maq}  by an additional factor $-8$. Other than this numerical discrepancy, our analysis confirm the main conclusion in~\cite{Kristiano-ml-2022maq}  that the one-loop corrections from small scale modes which leaves the horizon during the USR phase can induce significant correction to large scale power spectrum with an amplitude proportional to $e^{6 \Delta N}$.

One may wonder why our numerical prefactor  for $\Delta \calP$ is  different than the value obtained in~\cite{Kristiano-ml-2022maq}. We suspect that the assumption in~\cite{Kristiano-ml-2022maq} that the quartic Hamiltonian does not contribute to one-loop correction is unjustified which causes this discrepancy.   Indeed, we see non-trivial contributions from quartic Hamiltonian as studied in details in section~(\ref{quartic-H}). Having said this, one may try to compare our result for the cubic loop corrections, eq.~(\ref{cubic-power}), with the loop corrections from the cubic
interaction obtained in~\cite{Kristiano-ml-2022maq}. However, this comparison is somewhat misleading and inconclusive. The reason is that the cubic action in~\cite{Kristiano-ml-2022maq} is for $\calR$ (or in their notation $\zeta$) while our cubic Hamiltonian
${\bf H_3}$ is for $\pi$. It is important to note from eq.~(\ref{pi-R1}) that the relation between $\pi$ and $\calR$ is non-linear in the bulk. Therefore, the cubic action in~\cite{Kristiano-ml-2022maq}, once translated into our langue of using $\pi$,  receives additional contributions of order $\pi^4$ as well. Consequently, the cubic loop corrections in~\cite{Kristiano-ml-2022maq} would be different than our cubic loop correction eq.~(\ref{cubic-power}).  Therefore, it is misleading to compare our loop correction at the cubic order in $\pi$ with the corresponding cubic loop corrections in~\cite{Kristiano-ml-2022maq} obtained in terms of $\calR$. Only the \emph{total} one-loop corrections with the combined results from the cubic and quartic interactions
in our analysis can be compared with the \emph{total} one-loop corrections in~\cite{Kristiano-ml-2022maq}, including the suspected missing quartic interaction.

There are some important comments in order regarding our one-loop correction in power spectrum. The first comment is that in the extreme sharp transition with
$|h| \gg 1 $, the one-loop correction in power spectrum becomes arbitrarily large. To see this let us look at the fractional correction in power spectrum $\frac{\Delta \calP}{\calP_{\mathrm{CMB}}} $ from eq.~(\ref{loop-power}):
\begin{equation}
\label{fraction}
\frac{\Delta \calP}{\calP_{\mathrm{CMB}}} = \frac{6 \Delta N}{h}
   (  h^2  + 24 h + 180)  e^{6 \Delta N} \calP_{\mathrm{CMB}} \, .
\end{equation}
The above ratio depends on the two independent parameters $h$ and $\Delta N$.
If one takes the duration of the USR period to be long enough, or for a given value of $\Delta N$, takes $|h|$ to be arbitrarily large, then the above ratio approaches  unity, invalidating the perturbative analysis. For example,  for the standard sharp transition with $h=-6$   the ratio becomes order unity
for $\Delta N \simeq 2.3$. A longer period of USR phase will amplify  the amplitude  of curvature perturbations such that  the one-loop corrections become quickly larger than its tree-level value. Alternatively, one may  eliminate the parameter $h$ in the above ratio in favour of the final slow-roll parameter $\epsilon_0= \epsilon(\tau_0)$.
Using eq.~(\ref{ep-N}) (or directly from the definition of $h$ in eq.~(\ref{h-def}))
we have $| h| = 6 \sqrt\frac{\epsilon_0}{\epsilon_e}$  while $\epsilon_e = e^{-6 \Delta N} \epsilon_i$   so  in the limit of extreme sharp transition
$|h| \gg 1 $ we obtain
\begin{equation}
\bigg| \frac{\Delta \calP}{\calP_{\mathrm{CMB}}}  \bigg| \simeq
36   \sqrt\frac{\epsilon_0}{\epsilon_i}
(\Delta N e^{9 \Delta N}) \calP_{\mathrm{CMB}} \,  , \quad \quad
( |h| \gg 1 ) \, .
\end{equation}
If we further assume that there is no large hierarchy between the CMB scale $\epsilon_i$ and the final value $\epsilon_0$, we see that the above ratio scales like $36 \, e^{9 \Delta N} \calP_{\mathrm{CMB}}$. In this limit,  the above  ratio  becomes larger than unity when $\Delta N \simeq 1.9$
and the perturbative approximation breaks down. The conclusion is that  either by taking $|h|$ to be arbitrarily large or for a fixed value of  $|h|$  considering a long enough period of USR phase, one would invalidate the perturbation theory as the one-loop contribution surpasses the tree level contribution.

The second comment is the issue concerning PBHs formation as debated in~\cite{Kristiano-ml-2022maq, Riotto-ml-2023gpm, Riotto-ml-2023hoz}. As argued in~\cite{Kristiano-ml-2022maq} in order  to generate PBHs as candidate for dark matter from USR quantum fluctuations, one typically requires that for the USR modes
$\calP_{\mathrm{USR}} \sim 10^{-2}$ or so. However,  as discussed before, in order to trust the perturbative approximation we need $\frac{\Delta \calP}{\calP_{\mathrm{CMB}}}  \ll1$. Therefore, it is useful to combine these two constraints  to see if there is any window for the PBHs formation. The amplitude of curvature perturbations for the USR modes, measured at the end of inflation, is given by eq.~(\ref{PR-USR}). Combining this equation with the  fractional correction in power spectrum given in eq.~(\ref{fraction}) yields
\begin{equation}
 \calP_{\mathrm{USR}} = {\cal C}(\Delta N, h)  \bigg| \frac{\Delta \calP}{\calP_{\mathrm{CMB}}} \bigg| \, , \quad \quad
  {\cal C}(\Delta N, h) \equiv
\frac{( 6- h)^2}{ 6 |h| \Delta N  (  h^2  + 24 h + 180)} \, .
\end{equation}
In order to have a working mechanism of PBHs formation with
$\calP_{\mathrm{USR}} \sim 10^{-2}$ subject to the physical requirement
$\frac{\Delta \calP}{\calP_{\mathrm{CMB}}}  \ll 1$, we require  ${\cal C}$ no to be very small. For example, one may require ${\cal C} > 10^{-1}$ or so.
 Now let us estimate the numerical factor $ {\cal C}(\Delta N, h)$ for various values of $h$ with $\Delta N=2$.
 For $h=-6$,  as used in~\cite{Kristiano-ml-2022maq}, we obtain $ {\cal  C} \simeq  0.03  $ while for sharper transitions $h=-50$ and $h=-150$ we obtain $ {\cal C} \simeq  0.002  $ and  $ {\cal C} \simeq  0.0006$ respectively. It is clear that in the models with sharp transitions there is no practical chance of  PBHs formation as the numerical values of ${\cal C}$ are typically very small.

The third comment is that the contributions of the quartic and cubic loop corrections scale differently with $h$. It is instructive to look at their ratio, which from eqs.~(\ref{quartic-power}) and~(\ref{cubic-power}) is obtained to be
\begin{equation}
\label{ratio}
\frac{ \langle \calR_{\bfp_{1}} \calR_{\bfp_{2}} \rangle_{\Hb}}{ \langle \calR_{\bfp_{1}} \calR_{\bfp_{2}} \rangle_{\Ha}} = \frac{h^2 + 6h + 36}{18h + 144}  \, .
\end{equation}
For the sharp transitions with $|h| \gg1$, we see that the dominant loop correction comes from the quartic interaction. On the other hand, the case of $h=-6$ has the unique property that $\langle \calR_{\bfp_{1}} \calR_{\bfp_{2}} \rangle_{\bf{H_3}}= \langle \calR_{\bfp_{1}} \calR_{\bfp_{2}} \rangle_{\bf{H_4}}$. On the other hand, for $|h| <6$ the cubic Hamiltonian has the dominant contribution in the loop correction. Finally, the sign of both loop corrections is negative so there is no chance of cancellation between the cubic and quartic loop corrections.

The fourth comment is that eq.~(\ref{loop-power}) is obtained in the limit of sharp transition, corresponding to $h \leq -6$. A more realistic setup is when the transition from a USR phase to the final slow-roll phase happens mildly. As we discussed  in section~\ref{bispectrum}  in the setup of  two-phase inflation (USR$\rightarrow $ SR),   the effects of the transition on the final value of $f_{NL}$ were studied in details in~\cite{Cai-ml-2018dkf}.
It was demonstrated that the final value of $f_{NL}$ can be washed out if the transition is mild. The maximum value $f_{NL}=\frac{5}{2}$ happens only for the case of extreme sharp transition $|h| \gg 1 $ while for the standard  sharp transition with $h=-6$, one obtains a smaller value $f_{NL}=\frac{5}{8}$ as we demonstrated it explicitly  in section~\ref{bispectrum}. However, for a mild transition with  small value of $|h|$, the  $f_{NL}$ parameter falls off to the order of the slow-roll parameters (while the  Maldacena consistency condition is still violated). We believe that a similar effect can happen for the one-loop corrections in our three-phase
setup (SR$\rightarrow $ USR$\rightarrow $ SR) as well. If the transition takes place in a mild manner, then the leading order corrections proportional to $e^{6 \Delta N}$ will be washed out by the subsequent evolution of the mode function during the transition from the USR phase to the final slow-roll phase.
This point was advocated  in~\cite{Riotto-ml-2023gpm} as well. If so, the USR setup may still be employed as a viable mechanism to generate PBHs formation as a candidate for dark matter.

\section{Summary and discussions }
\label{Summary}

In this work we have revisited the question of one-loop effects on large scale curvature perturbations power spectrum induced from small scale modes which
become superhorizon during an intermediate USR phase. This is an important question both theoretically and also for its implications for the PBHs formation.

Our results largely support the conclusion made in~\cite{Kristiano-ml-2022maq, Kristiano-ml-2023scm} that the induced one-loop corrections can become large enough to kill the popular mechanism of PBHs formation in the USR setup. As
the one-loop corrections become very large they invalidate the perturbative assumption
before one proceeds to calculate the PBHs abundance in this setup.
Having said this, there is a discrepancy between our results and those of~\cite{Kristiano-ml-2022maq, Kristiano-ml-2023scm} as our result is larger by a factor
of $8$ compared to that of~\cite{Kristiano-ml-2022maq, Kristiano-ml-2023scm} in the limit where these two calculations have overlap (i.e.\ in the special case $h=-6$).
 We believe this discrepancy is caused by the assumption in~\cite{Kristiano-ml-2022maq, Kristiano-ml-2023scm}
that the quartic self interaction does not contribute to the one-loop calculation. Our direct calculation based on EFT formalism  shows that there are non-trivial contributions from the quartic Hamiltonian which yield to a comparable contribution to one-loop power spectrum as does the quartic Hamiltonian.

Our analytical results were obtained in the limit of sharp transition where one neglects the relaxation period after the USR phase. An interesting conclusion
from our result is that for an infinitely sharp transition the induced one-loop correction can be arbitrarily large, invalidating the perturbative approximation completely. Of course,
in a realistic case, one needs a relaxation period so it is possible that much of the one-loop corrections are washed out during the subsequent evolution of the mode function  after the USR phase. This was certainly the case in $f_{NL}$ story as demonstrated explicitly in~\cite{Cai-ml-2018dkf}. This will also support the arguments in~\cite{Riotto-ml-2023gpm, Riotto-ml-2023hoz} that in a realistic setup with a mild transition, the one-loop corrections become harmless
and one can still employ the USR setup for the PBHs formation. We would like to come back to this important question and examine the effects of a mild transition on the one-loop calculations.

\looseness=-1
As usual in QFT analysis, in performing the loop corrections one has to take into account the renormalization of the divergences, both UV and IR. In this work, we were mainly interested in the contributions from the modes in the interval  $q_i \leq q \leq q_e$ which become superhorizon during the USR phase. Our main goal was to check if there is a cancellation between the cubic and quartic contributions for the one-loop corrections. Also our second main goal was to investigate the roles of the sharpness of the transition. For these motivations the renormalization of the power spectrum  is not an immediate concern. However, to read off the final physical power spectrum  one has to examine the divergence associated to the  modes with momentum larger than $q_e$ and take into account the question of renormalization.

There are a number of directions in which the current analysis can be extended.
One interesting direction to proceed is to consider the effects of a non-trivial sound speed $c_s$ for the induced one-loop corrections in power spectrum. As a related comment we mention that the relation between the one-loop correction in power spectrum and the amplitude of non-Gaussianity was studied in~\cite{Kristiano-ml-2021urj}. It was concluded that in order for the one-loop corrections in a model of inflation with small  $c_s$ to be under control, then the amplitude of the equilateral-type non-Gaussianity has to be small.   The other interesting question
is to calculate the one-loop correction using the formalism of stochastic inflation. Indeed, this question was already studied in~\cite{Firouzjahi-ml-2018vet, Firouzjahi-ml-2020jrj} in the USR setup. It was shown there that $\Delta \calP \propto \calP^2 \propto e^{12 \Delta N}$. This is unlike our current case where $\Delta \calP \propto e^{6 \Delta N} $. The reason is that in~\cite{Firouzjahi-ml-2018vet, Firouzjahi-ml-2020jrj} one has a two-phase model where an early USR phase is followed by a slow-roll phase and the CMB scale modes leave the horizon during the USR phase. However, in our current analysis, we have a three-phase setup of inflation in which the CMB scale modes leave the horizon at the first slow-roll stage long prior to the USR phase. Finally, the other interesting question is to investigate the one-loop corrections  for the tensor power spectrum. As there are interactions between two scalars and one tensor perturbations, then naturally there can be large one-loop corrections in tensor power spectrum as well. We would like to come back to these interesting questions in future.

\acknowledgments  I am grateful to   Mohammad Hossein Namjoo for many insightful discussions during the progress of this work. I would like to  thank
 Xingang Chen, Junichi Yokoyama, Mahdiyar Noorbala, Jason Kristiano and Sina Hooshangi for useful  discussions and comments on the draft.  I also thank Amin Nassiri-Rad and Kosar Asadi for checking the analysis in the revised draft
 and for pointing out the typos.  We thank the anonymous referee for the insightful comments and for pointing out corrections.
 This research is  partially supported by the ``Saramadan'' Federation of Iran.

\appendix
\section{Loop corrections from the cubic Hamiltonian }
 \label{appendix}

 In this appendix, we present the details of the analysis for the loop corrections from
 the cubic Hamiltonian.

 As mentioned in section~\ref{cubic-H},   expanding the Dyson series to second order in $\Ha$ we have
\begin{equation}
   \langle \calR_{\bfp_{1}}(\tau_0) \calR_{\bfp_{2}}(\tau_0) \rangle_{\Ha} =    \langle \calR_{\bfp_{1}}(\tau_0) \calR_{\bfp_{2}}(\tau_0) \rangle_{(2,0)} +   \langle \calR_{\bfp_{1}}(\tau_0) \calR_{\bfp_{2}}(\tau_0) \rangle_{(1,1)} +    \langle \calR_{\bfp_{1}}(\tau_0) \calR_{\bfp_{2}}(\tau_0) \rangle_{(0, 2)}
  \end{equation}
  in which
\begin{eqnarray}
  \label{20-int-b}
\langle \calR_{\bfp_{1}}(\tau_0) \calR_{\bfp_{2}}(\tau_0) \rangle_{(2,0)} &=&
- \int_{-\infty}^{\tau_0} d \tau_1 \int_{-\infty}^{\tau_1} d \tau_2
\big \langle \Ha (\tau_2)  \Ha (\tau_1) \calR_{\bfp_{1}}(\tau_0) \calR_{\bfp_{2}}(\tau_0)
 \big \rangle   \nonumber\\
&=& \langle \calR_{\bfp_{1}}(\tau_0) \calR_{\bfp_{2}}(\tau_0)
\rangle^\dagger_{(0,2)}\, ,
\end{eqnarray}
and
 \begin{equation}
  \label{11-int-b}
\langle \calR_{\bfp_{1}}(\tau_0) \calR_{\bfp_{2}}(\tau_0) \rangle_{(1,1)} =
 \int_{-\infty}^{\tau_0} d \tau_1 \int_{-\infty}^{\tau_0} d \tau_2
\big \langle \Ha (\tau_1)   \calR_{\bfp_{1}}(\tau_0) \calR_{\bfp_{2}}(\tau_0)
\Ha (\tau_2) \big \rangle   \, .
\end{equation}

As explained in the main  text, the dominant contributions for the time integrals above  come from the USR phase as the mode functions are growing and the time derivatives are not slow-roll suppressed. During the USR phase $\eta=-6$ so it simply can be taken out of the time integrals. At this stage, we consider only the interactions containing the time derivative, $\pi \pi'^2$ and implement the contributions of the gradient interactions $\pi (\partial \pi)^2$ at the end.

With this discussion in mind, let us start with eq.~(\ref{11-int-b}).
After performing all the contractions and considering the limit that $| \bfp| \ll |\bfq|$, we obtain four different contributions in $\langle \calR_{\bfp_{1}}(\tau_0) \calR_{\bfp_{2}}(\tau_0) \rangle_{(1,1)}$ denoted by
$A_i$ for $i=1\ldots 4$. For example, $A_1$ is given by
 \begin{equation}
  \label{A1-aa}
A_1 =  4 \eta^2  M_P^4 \int \frac{d^3 \bfq}{(2 \pi)^3}
 \int_{\tau_i}^{\tau_e} d \tau_1 \epsilon a^2 \int_{\tau_i}^{\tau_e} d \tau_2 \epsilon a^2
\calR_p(\tau_e)  \calR^*_p(\tau_e)  \calR_p(\tau_1)  \calR^*_p(\tau_2)
\calR'_q(\tau_1)^2 \calR^{'*}_q(\tau_2)^2   \, ,
\end{equation}
with similar expressions for $A_2, A_3$ and $A_4$.

Defining
\begin{equation}
\label{X1-def}
 X_1(\tau) \equiv  \eta \epsilon a^2  \calR_p^*(\tau_0)  \calR_p(\tau) \calR'_q(\tau)^2 \, ,
 \end{equation}
 we can express $A_1$ as follows
 \begin{equation}
\label{A1}
A_1=  4   M_P^4 \int \frac{d^3 \bfq}{(2 \pi)^3}
 \int_{\tau_i}^{\tau_e} d \tau_1X_1 (\tau_1)  \int_{\tau_i}^{\tau_e} d \tau_2
 X_1^*(\tau_2) \, .
 \end{equation}
 Similarly, defining $X_2$ as
 \begin{equation}
 \label{X2-def}
 X_2(\tau) \equiv  \eta \epsilon a^2  \calR_p^*(\tau_0)  \calR'_p(\tau) \calR'_q(\tau)
 \calR_q(\tau) \, ,
 \end{equation}
 with some efforts one can show that the remaining terms $A_1, A_2$ an $A_3$ can be written as follows
 \begin{align}
 A_2&=  8   M_P^4 \int \frac{d^3 \bfq}{(2 \pi)^3}
 \int_{\tau_i}^{\tau_e} d \tau_1X_1 (\tau_1)  \int_{\tau_i}^{\tau_e} d \tau_2
 X_2^*(\tau_2)  \, ,
 \\
 A_3&=  8   M_P^4 \int \frac{d^3 \bfq}{(2 \pi)^3}
 \int_{\tau_i}^{\tau_e} d \tau_1X_1^* (\tau_1)  \int_{\tau_i}^{\tau_e} d \tau_2
 X_2(\tau_2)  \, ,
 \end{align}
 and
 \begin{equation}
 A_4=  16   M_P^4 \int \frac{d^3 \bfq}{(2 \pi)^3}
 \int_{\tau_i}^{\tau_e} d \tau_1X_2 (\tau_1)  \int_{\tau_i}^{\tau_e} d \tau_2
 X_2^*(\tau_2) \, .
 \end{equation}

Using the asymptotic form of the mode function at $\tau=\tau_0$ one can check that
 $X_1 \propto p^{-3}$  with the subleading correction of order $p^{-1}$ while
 $X_2 \propto p^{-1}$ with the subleading correction of order $p^{0}$. Since we have the large hierarchy $p \ll q$, we can safely neglect the contribution of $A_4$ in the following analysis. Intuitively speaking, $X_2$ has a time derivative $\calR'_p(\tau)$
 compared to $X_1$ which has no corresponding time derivative so naturally $X_2$ is further suppressed compared to $X_1$ by an extra factor of $p$.

Now we calculate   $\langle \calR_{\bfp_{1}}(\tau_0) \calR_{\bfp_{2}}(\tau_0) \rangle_{(2,0)} $ again considering only the time derivative interaction $\pi \pi'^2$.
Performing all the contractions, we have four different contributions
denoted by $B_i, i=1,\ldots 4$. For example, $B_1$ has the following form
 \begin{equation}
  \label{A1-a}
B_1 =  -4 \eta^2  M_P^4 \int \frac{d^3 \bfq}{(2 \pi)^3}
 \int_{\tau_i}^{\tau_e} d \tau_1 \epsilon a^2 \int_{\tau_i}^{\tau_1} d \tau_2 \epsilon a^2 \calR^*_p(\tau_0)^2  \calR_p(\tau_1)  \calR_p(\tau_2)
\calR'_q(\tau_2)^2 \calR^{'*}_q(\tau_1)^2   ,
\end{equation}
with similar expressions for $B_2, B_3$ and $B_4$.

To calculate $B_i$,  we further define the quantities $Z_1$ and $Z_2$ as follows:
\begin{equation}
\label{Z1-def}
 Z_1(\tau) \equiv  \eta \epsilon a^2  \calR_p^*(\tau_0)  \calR_p(\tau) \calR^{'*}_q(\tau)^2 \, ,
 \end{equation}
 and
\begin{equation}
 \label{Z2-def}
 Z_2(\tau) \equiv  \eta \epsilon a^2  \calR_p^*(\tau_0)  \calR'_p(\tau) \calR^{'*}_q(\tau)
 \calR^*_q(\tau) \, .
 \end{equation}
We note that in their dependence on $p$, $Z_1$ is similar to $X_1$ while $Z_2$ is similar to $X_2$.  With these definition, one obtains
\begin{align}
B_1 &= -4   M_P^4 \int \frac{d^3 \bfq}{(2 \pi)^3}
 \int_{\tau_i}^{\tau_e} d \tau_1Z_1 (\tau_1)  \int_{\tau_i}^{\tau_1} d \tau_2
 X_1(\tau_2) \, ,
\\
B_2 &= -8   M_P^4 \int \frac{d^3 \bfq}{(2 \pi)^3}
 \int_{\tau_i}^{\tau_e} d \tau_1Z_2 (\tau_1)  \int_{\tau_i}^{\tau_1} d \tau_2
 X_1(\tau_2) \, ,
\\
B_3 &= -8   M_P^4 \int \frac{d^3 \bfq}{(2 \pi)^3}
 \int_{\tau_i}^{\tau_e} d \tau_1Z_1 (\tau_1)  \int_{\tau_i}^{\tau_1} d \tau_2
 X_2(\tau_2) \, ,
\end{align}
and
\begin{equation}
B_4 = -16   M_P^4 \int \frac{d^3 \bfq}{(2 \pi)^3}
 \int_{\tau_i}^{\tau_e} d \tau_1 Z_2 (\tau_1)  \int_{\tau_i}^{\tau_1} d \tau_2
 X_2 (\tau_2) \, .
\end{equation}

To proceed further, we note that
\begin{equation}
Z_1= X_1^* + i \delta^*
\end{equation}
in which
\begin{equation}
 \delta(\tau)  \equiv 2 \epsilon \eta a^2 \calR'_q(\tau)^2
 \mathrm{Im}  \big[  \calR_p^*(\tau_0)  \calR_p(\tau)   \big]  \, ,
\end{equation}
and
\begin{equation}
Z_2= X_2^* + i \beta^*
\end{equation}
in which
\begin{equation}
 \beta (\tau) \equiv 2 \epsilon \eta a^2 \calR'_q(\tau) \calR_q(\tau)
 \mathrm{Im}
 \big[  \calR_p^*(\tau_0)  \calR'_p(\tau)   \big] \, .
 \end{equation}

Combining the results for $ \langle \calR_{\bfp_{1}} \calR_{\bfp_{2}} \rangle_{(2,0)}= \langle \calR_{\bfp_{1}} \calR_{\bfp_{2}} \rangle_{(0, 2)}^\dagger $ and
$  \langle \calR_{\bfp_{1}} \calR_{\bfp_{2}} \rangle_{(1,1)} $ and after some algebra one obtains
\begin{equation}
\label{H3-int}
   \langle \calR_{\bfp_{1}}(\tau_0) \calR_{\bfp_{2}}(\tau_0) \rangle_{\pi \pi'^2} =
   - 8 M_P^4  \int \frac{d^3 \bfq}{(2 \pi)^3}  \int_{\tau_i}^{\tau_e} d \tau_1
    \int_{\tau_i}^{\tau_1} d \tau_2 \,   \mathrm{Im} \Big[ X^*_1( \tau_2)
    \big( \delta (\tau_1) + 2 \beta ( \tau_1)\,  \big)  \Big] \, .
\end{equation}
Note that in obtaining the above results we have neglected the contributions from the  terms $A_4$ and $B_4$ which are subleading as mentioned previously.

We can simplify  $\delta$ and $\beta$ in the integral~(\ref{H3-int}) to some extent.  Using  the asymptotic form of the mode function,  we have:
\begin{equation}
 \mathrm{Im}  \big[  \calR_p^*(\tau_0)  \calR_p(\tau)   \big]  = \frac{H^2 \tau_i^6}{12 M_P^2 \epsilon_i  h \tau_e^3 \tau^3 } \big( h \tau_e^3 + (6- h) \tau^3
 \big) \, ,
\end{equation}
and
\begin{equation}
 \mathrm{Im}  \big[  \calR_p^*(\tau_0)  \calR'_p(\tau)   \big]  = -\frac{H^2 \tau_i^6}{4 M_P^2 \epsilon_i   \tau^4 }  \, .
\end{equation}
Using these expressions in $\delta$ and $\beta$ and plugging them in~(\ref{H3-int}) we obtain\footnote{To perform the integrals in eqs.~(\ref{H3-int}) and~(\ref{H3-int2}) we employ the Maple computational software. }
\begin{equation}
\label{H3-int2}
   \langle \calR_{\bfp_{1}}(\tau_0) \calR_{\bfp_{2}}(\tau_0) \rangle_{\pi \pi'^2} =
   \frac{ 9 (5 h + 36)}{8h}
   \big( \Delta N e^{6 \Delta N} \big)
 \frac{H^4  }{  2 \pi^2 M_P^4 \epsilon_i^2 p^3 }
  \end{equation}
  To calculate the nested in-in integrals, we have chosen the strategy to integrate only over the modes which become superhorizon during the USR phase. For a given momentum $q$ running in the loop, we consider $-\frac{1}{q} \leq \tau_2 \leq \tau_1\leq \tau_e$. Alternatively, one can consider a different strategy and integrate over   the entire range $\tau_i \leq \tau_2 \leq \tau_1\leq \tau_e$ independent of the value of $q$. This only changes the numerical factors in Eq. (\ref{H3-int2}).

So far we have considered the Hamiltonian containing the time derivative $\pi \pi'^2$. Now let us consider the remaining part of the Hamiltonian involving the gradient term $\pi (\partial \pi)^2$. It is convenient to use ${\bf H_3}$ as given in the second line of Eq. (\ref{H3}). In dealing with the gradient terms there are two different types of contributions in the nested integrals.  The first type is the cross-contribution, one source from  $\pi \pi'^2$ interaction and the other source from 
$\pi^2 \partial^2 \pi$. The second type  is when both interactions in the nested integral are sourced by $\pi^2 \partial^2 \pi$. With some efforts one can show that the second type, involving two powers of $\pi^2 \partial^2 \pi$ in the 
nested integrals, is subleading. This is easy to understand because  we are mostly interested in the regime where $q |\tau| <1 $.  

Performing the contractions and following a similar steps as outlined in the above analysis, and after a long and tedious calculation,  one obtains  
\begin{equation}
\label{cubic-rsult2}
   \langle \calR_{\bfp_{1}}(\tau_0) \calR_{\bfp_{2}}(\tau_0)  \rangle_{\Ha} =
     - 8 M_P^4  \int \frac{d^3 \bfq}{(2 \pi)^3}
     \int_{\tau_i}^{\tau_e} d \tau_1
    \int_{\tau_i}^{\tau_1} d \tau_2 \,  {\cal F} (\tau_1, \tau_2; q) \, ,
         \end{equation}
 with ${\cal F} (\tau_1, \tau_2; q)$  given by
  \begin{equation}
  \label{F-def}
  {\cal F} (\tau_1, \tau_2; q) \equiv  \mathrm{Im} \left\{ X^*_1(  \tau_2) \Big[  \Big( 1- f^*_q(\tau_2)  \Big) \Big( \, \delta (\tau_1) + 2 \beta ( \tau_1)\,  \Big)     
  - f_q(\tau_1) \delta (\tau_1)  \Big] \right\} \, ,
 \end{equation}
 in which 
 \begin{equation}
f_q(\tau)  \equiv \frac{\big( \partial \calR_q \big)^2}{  \calR_q'^2} = \frac{q^2 \calR_q^2}{\calR_q'^2}. 
\end{equation}
 In the above expression, the factor $f_q(\tau)$ captures the effects from the gradient interactions $\pi^2 \partial^2 \pi$. 
 Note that if we set $f_q(\tau_i)=0$ in eq.~(\ref{F-def}), then we obtain the result in eq.~(\ref{H3-int}) for the $\pi \pi'^2$ interaction.

After performing the nested integral, it turns out that the contribution from the last term in  eq.~(\ref{F-def}) (containing $f_q(\tau_1) \delta (\tau_1)$) 
is subleading compared to the first two terms. The final result is obtained to be 
 \begin{equation}
\label{cubic-rsult3}
   \langle \calR_{\bfp_{1}}(\tau_0) \calR_{\bfp_{2}}(\tau_0)  \rangle_{\Ha} =
    \frac{ 27 ( h + 8)}{4h}
   \big( \Delta N e^{6 \Delta N} \big)
 \frac{H^4  }{  2 \pi^2 M_P^4 \epsilon_i^2 p^3 }  \, .
   \end{equation}

As discussed previously, in obtaining the above result we have imposed the requirement  that the mode to be superhorizon during the USR phase so for a given $q$, we take $-\frac{1}{q} \leq \tau_2 \leq \tau_1\leq \tau_e$. Alternatively, one may drop  this requirement and, independent of the value of  $q$, integrate over the whole range $\tau_i \leq \tau_2 \leq \tau_1\leq \tau_e$. Performing the nested integral using this prescription modifies the prefactor 
in Eq. (\ref{cubic-rsult3}) such that 
 \begin{equation}
\label{cubic-rsult3b}
   \langle \calR_{\bfp_{1}}(\tau_0) \calR_{\bfp_{2}}(\tau_0)  \rangle_{\Ha} =
    \frac{9 ( h -12)}{8h}
   \big( \Delta N e^{6 \Delta N} \big)
 \frac{H^4  }{  2 \pi^2 M_P^4 \epsilon_i^2 p^3 }  \quad \quad 
 \mathrm{(second \, \, strategy)}  \, .
 \end{equation}
%Having said this, we believe the first strategy is more physical since only the modes which become superhorizon during the USR phase are expected to affect the long mode. However, this does  not change the result significantly  as  both Eqs. (\ref{cubic-rsult3}) and (\ref{cubic-rsult3b}) qualitatively look very similar. 

\end{document}